\documentclass[useAMS,usenatbib,usegraphicx]{mn2e}
\renewcommand{\th}   {$T_{\rm h}$}                 \newcommand{\Th}     {T_{\rm h}}                 
\newcommand{\eh}     {$E_{\rm hot}$}               \newcommand{\Eh}     {E_{\rm hot}}             
\newcommand{\tc}     {$T_{\rm c}$}                 \newcommand{\Tc}     {T_{\rm c}}               
\newcommand{\nh}     {$n_{\rm h}$}                 \newcommand{\Nh}     {n_{\rm h}}               
\newcommand{\nc}     {$n_{\rm c}$}                 \newcommand{\Nc}     {n_{\rm c}}               
\newcommand{\mhot}   {$M_{\rm hot}$}               \newcommand{\Mhot}   {M_{\rm hot}}		  
\newcommand{\mcold}  {$M_{\rm cold}$}              \newcommand{\Mcold}  {M_{\rm cold}}            
\newcommand{\mstar}  {$M_\star$}                                    
\newcommand{\mhalo}  {$M_{\rm halo}$}              \newcommand{\Mhalo}  {M_{\rm halo}}            
\newcommand{\mtot}   {$M_{\rm tot}$}

               \newcommand{\Zhot}   {Z_{\rm hot}}             
\newcommand{\zcold}  {$Z_{\rm cold}$}              \newcommand{\Zcold}  {Z_{\rm cold}}            
\newcommand{\zstar}  {$Z_\star$}                                    
              \newcommand{\Zhalo}  {Z_{\rm halo}}            
\newcommand{\muh}    {$\mu_{\rm h}$}               \newcommand{\Muh}    {\mu_{\rm h}}             
\newcommand{\muc}    {$\mu_{\rm c}$}               \newcommand{\Muc}    {\mu_{\rm c}}             
\renewcommand{\fh}   {$f_{\rm h}$}                 \newcommand{\Fh}     {f_{\rm h}}               
\newcommand{\fc}     {$f_{\rm c}$}                 \newcommand{\Fc}     {f_{\rm c}}               
\newcommand{\shape}  {$\mu_{\rm shape}$}           \newcommand{\Shape}  {\mu_{\rm shape}}         
\newcommand{\fcoll}  {$f_{\rm coll}$}              \newcommand{\Fcoll}  {f_{\rm coll}}            
\newcommand{\fpds}   {$f_{\rm pds}$}               \newcommand{\Fpds}   {f_{\rm pds}}             
\newcommand{\fcool}  {$f_{\rm cool}$}              \newcommand{\Fcool}  {f_{\rm cool}}            
\newcommand{\fstar}  {$f_\star$}                   \newcommand{\Fstar}  {f_\star}                 
\newcommand{\fbom}   {$f_{\rm bo,max}$}            	  
\newcommand{\fevap}  {$f_{\rm evap}$}              \newcommand{\Fevap}  {f_{\rm evap}}            
\newcommand{\tevap}  {$T_{\rm evap}$}              \newcommand{\Tevap}  {T_{\rm evap}}            
\newcommand{\tff}    {$t_{\rm dyn}$}               \newcommand{\Tff}    {t_{\rm dyn}}             
\newcommand{\mstsn}  {$M_{\star{\rm ,sn}}$}        \newcommand{\Mstsn}  {M_{\star{\rm ,sn}}}
\newcommand{\tlife}  {$t_{\rm life}$}              \newcommand{\Tlife}  {t_{\rm life}}            
\newcommand{\mcc}    {$m_{\rm cc}$}                \newcommand{\Mcc}    {m_{\rm cc}}              
\newcommand{\ncc}    {$N_{\rm cc}$}                \newcommand{\Ncc}    {N_{\rm cc}}              
\newcommand{\nsn}    {$N_{\rm sn}$}                \newcommand{\Nsn}    {N_{\rm sn}}              
\newcommand{\rsn}    {$R_{\rm sn}$}                \newcommand{\Rsn}    {R_{\rm sn}}              
     \newcommand{\Rsbad}  {R_{\rm sb}^{\rm (ad)}}     
    \newcommand{\Rsbpds} {R_{\rm sb}^{\rm (pds)}}    
\newcommand{\eth}    {$E_{\rm sb}^{\rm (th)}$}     \newcommand{\Eth}    {E_{\rm sb}^{\rm (th)}}	  
\newcommand{\ekin}   {$E_{\rm sb}^{\rm (kin)}$}    \newcommand{\Ekin}   {E_{\rm sb}^{\rm (kin)}}  
\newcommand{\tpds}   {$t_{\rm pds}$}               \newcommand{\Tpds}   {t_{\rm pds}}             
\newcommand{\rpds}   {$R_{\rm pds}$}                            
\newcommand{\heff}   {$H_{\rm eff}$}               \newcommand{\Heff}   {H_{\rm eff}}             
                        
\newcommand{\heffh}  {$H_{\rm eff,h}$}                        
\newcommand{\hfin}   {$H_{\rm h}$}                 \newcommand{\Hfin}   {H_{\rm h}}               
\newcommand{\fbo}    {$f_{\rm bo}$}                \newcommand{\Fbo}    {f_{\rm bo}}              
\newcommand{\tbo}    {$t_{\rm bo}$}                \newcommand{\Tbo}    {t_{\rm bo}}              
\newcommand{\rbos}   {$R_{\rm bo'}$}                            
\newcommand{\tbos}   {$t_{\rm bo'}$}               \newcommand{\Tbos}   {t_{\rm bo'}}             
\newcommand{\rfin}   {$R_{\rm fin}$}               \newcommand{\Rfin}   {R_{\rm fin}}             
   \newcommand{\Tconfad}{t_{\rm conf}^{\rm (ad)}}
  
\newcommand{\tconf}  {$t_{\rm conf}$}              \newcommand{\Tconf}  {t_{\rm conf}}            
\newcommand{\tfin}   {$t_{\rm fin}$}               \newcommand{\Tfin}   {t_{\rm fin}}

\newcommand{\rconf}  {$R_{\rm conf}$}                          
\newcommand{\tcool}  {$t_{\rm cool}$}              \newcommand{\Tcool}  {t_{\rm cool}} 		  
\newcommand{\tinf}   {$t_{\rm inf}$}               \newcommand{\Tinf}   {t_{\rm inf}} 		  
\newcommand{\tsf}    {$t_{\rm sf}$}                \newcommand{\Tsf}    {t_{\rm sf}}		  
\newcommand{\poro}   {$Q_{\rm sb}$}                \newcommand{\Poro}   {Q_{\rm sb}}              
         \newcommand{\Dmhot}  {\dot{M}_{\rm hot}}       
        \newcommand{\Dmcold} {\dot{M}_{\rm cold}}      
             \newcommand{\Dmstar} {\dot{M}_\star}           
        \newcommand{\Dmhalo} {\dot{M}_{\rm halo}}      
\newcommand{\dmcool} {$\dot{M}_{\rm cool}$}        \newcommand{\Dmcool} {\dot{M}_{\rm cool}}      
\newcommand{\dmsnpl} {$\dot{M}_{\rm snpl}$}        \newcommand{\Dmsnpl} {\dot{M}_{\rm snpl}}      
\newcommand{\dmev}   {$\dot{M}_{\rm evap}$}        \newcommand{\Dmev}   {\dot{M}_{\rm evap}}      
          \newcommand{\Dmbo}   {\dot{M}_{\rm bo}}        
\newcommand{\dmrest} {$\dot{M}_{\rm rest}$}        \newcommand{\Dmrest} {\dot{M}_{\rm rest}}      
\newcommand{\dmint}  {$\dot{M}_{\rm int}$}         \newcommand{\Dmint}  {\dot{M}_{\rm int}}       
\newcommand{\dmsfr}  {$\dot{M}_{\rm sf}$}          \newcommand{\Dmsfr}  {\dot{M}_{\rm sf}}       
\newcommand{\dminf}  {$\dot{M}_{\rm inf}$}         \newcommand{\Dminf}  {\dot{M}_{\rm inf}}       
\newcommand{\tleak}  {$t_{\rm leak}$}              \newcommand{\Tleak}  {t_{\rm leak}}		  
\newcommand{\dmleak} {$\dot{M}_{\rm leak}$}        \newcommand{\Dmleak} {\dot{M}_{\rm leak}}      
\newcommand{\frest}  {$f_{\rm rest}$}              \newcommand{\Frest}  {f_{\rm rest}}            
\newcommand{\desn}   {$\dot{E}_{\rm sn}$}          \newcommand{\Desn}   {\dot{E}_{\rm sn}}	  
\newcommand{\decool} {$\dot{E}_{\rm cool}$}        \newcommand{\Decool} {\dot{E}_{\rm cool}}	  
\newcommand{\desnpl} {$\dot{E}_{\rm snpl}$}        \newcommand{\Desnpl} {\dot{E}_{\rm snpl}}	  
\newcommand{\debo}   {$\dot{E}_{\rm bo}$}          \newcommand{\Debo}   {\dot{E}_{\rm bo}}	  
\newcommand{\desb}   {$\dot{E}_{\rm fb}$}          \newcommand{\Desb}   {\dot{E}_{\rm fb}}	  
\newcommand{\dehot}  {$\dot{E}_{\rm hot}$}         \newcommand{\Dehot}  {\dot{E}_{\rm hot}}	  
\newcommand{\deleak} {$\dot{E}_{\rm leak}$}        \newcommand{\Deleak} {\dot{E}_{\rm leak}}      
\newcommand{\fe}     {$f_{\rm E}$}                 \newcommand{\Fe}     {f_{\rm E}}		  
\newcommand{\ro}     {$\rho_{\rm tot}$}            	  
\newcommand{\stot}   {$\Sigma_{\rm tot}$}          \newcommand{\Stot}   {\Sigma_{\rm tot}}        
\newcommand{\esn}{$E_{51}$}
\newcommand{\lmech}{$L_{38}$}
\newcommand{\circa}{$\sim$}
\newcommand{\K}{K}
\newcommand{\press}{K cm$^{-3}$}
\newcommand{\yr}{yr}
\newcommand{\Gyr}{Gyr}
\newcommand{\pc}{pc}
\newcommand{\kpc}{kpc}
\newcommand{\kms}{km s$^{-1}$}
\newcommand{\cmt}{cm$^{-3}$}
\newcommand{\msun}{M$_\odot$}
\newcommand{\msunyr}{M$_\odot$ yr$^{-1}$}

\newcommand{\dens}{M$_\odot$ pc$^{-3}$}
\newcommand{\surf}{M$_\odot$ pc$^{-2}$}
\newcommand{\be}{\begin{equation}}
\newcommand{\ee}{\end{equation}}
\newcommand{\cl}{_{\rm cl}}
\newcommand{\alfa}{$\alpha_{\rm cl}$}
\newcommand{\sigv}{$\sigma_{\rm v}$}

\newcommand{\roh}{$\bar{\rho}_{\rm h}$}
\newcommand{\roc}{$\bar{\rho}_{\rm c}$}
\newcommand{\uno}{$M_{\rm i}$}         
\newcommand{\due}{$M^{Z}_{\rm i}$}     
\newcommand{\tre}{$\dot{M}_{\rm i}$}
\newcommand{\qua}{$\dot{M}^{Z}_{\rm i}$}
\newcommand{\tcross}{$t_{\rm cross}$}
\newcommand{\tcg}{$t_{\rm coag}$}

\title[Physical regimes for feedback]
{Physical regimes for feedback in galaxy formation}
\author[P. Monaco]{Pierluigi Monaco\\
Dipartimento di Astronomia, Universit\`a di Trieste, 
via Tiepolo 11, 34131 Trieste, Italy - email: monaco@ts.astro.it}

\begin{document}

\date{Accepted ... Received ...}

\pagerange{\pageref{firstpage}--\pageref{lastpage}} \pubyear{2003}

\maketitle

\label{firstpage}

\begin{abstract}

We present a new (semi-)analytic model for feedback in galaxy
formation.  The interstellar medium (hereafter ISM) is modeled as a
two-phase medium in pressure equilibrium, where the cold phase is
fragmented into clouds with a given mass spectrum.  Cold gas infalls
from an external halo.  Large clouds are continually formed by
coagulation and destroyed by gravitational collapse.  Stars form in
the collapsing clouds; the remnants of exploding type II supernovae
(hereafter SNe) percolate into a single super-bubble (hereafter SB)
that sweeps the ISM, heating the hot phase (if the SB is adiabatic) or
cooling it (in the snowplow stage, when the interior gas of the SB has
cooled).  Different feedback regimes are obtained whenever SBs are
stopped either in the adiabatic or in the snowplow stage, either by
pressure confinement or by blow-out.

The resulting feedback regimes occur in well-defined regions of the
space defined by vertical scale-length and surface density of the
structure.  In the adiabatic blow-out regime the efficiency of SNe in
heating the ISM is rather low (\circa5 per cent, with \circa80 per
cent of the energy budget injected into the external halo), and the
outcoming ISM is self-regulated to a state that, in conditions typical
of our galaxy, is similar to that found in the Milky Way.  Feedback is
most efficient in the adiabatic confinement regime, where
star-formation is hampered by the very high thermal pressure and the
resulting inefficient coagulation.  In some significant regions of the
parameter space confinement takes place in the snowplow stage; in this
case the hot phase has a lower temperature and star formation is
quicker.  In some critical cases, found at different densities in
several regions of the parameter space, the hot phase is strongly
depleted and the cold phase percolates the whole volume, giving rise
to a burst of star formation.

While the hot phase is allowed to leak out of the star-forming region,
and may give rise to a tenuous wind that escapes the potential well of
a small galactic halo, strong galactic winds are predicted to happen
only in critical cases or in the snowplow confinement regime whenever
the SBs are able to percolate the volume.

This model provides a starting point for constructing a realistic grid
of feedback solutions to be used in galaxy formation codes, either
semi-analytic or numeric.  The predictive power of this model extends
to many properties of the ISM, so that most parameters can be
constrained by reproducing the main properties of the Milky Way.

\end{abstract}

\begin{keywords}
galaxies: formation -- galaxies: ISM -- ISM: bubbles -- ISM:
kinematics and dynamics
\end{keywords}

\section{Introduction}

Galaxy formation is an open problem.  This is due to the complexity of
the feedback processes that arise from the energetic activity of
massive or dying stars, taking place through winds, ionizing photons
and SN explosions (not to mention AGN).  These feedback processes
involve a large range of scales and masses, from the sub-pc scale of
star formation to the $\ga$10 \kpc\ scale of galactic winds, and from
1 to $10^{12}$ \msun\ or more.

It is useful at this stage to identify ranges of scales in which
different processes are dominant.  On $\ga$1 \kpc\ spatial and
$\ga$$10^6$ \msun\ mass scales the dominant processes such as shock
heating of gas, radiative cooling, disc formation, galaxy merging and
tidal or ram-pressure stripping are closely related to the dark-matter
halo hosting the galaxy and to its hierarchical assembly.  On scales
ranging from $\sim$1 \pc\ to $\sim$1 \kpc, or from $\sim$1000 to
$\sim$$10^6$ \msun, cool gas reaches suitable conditions for collapse
and star formation, and the energy input from massive stars (through
winds, UV photons and SNe) acts in shaping and sustaining the
multi-phase structure of the ISM.  At smaller scales star formation
takes place; it is most likely driven and self-limited by
magneto-hydro-dynamical (MHD) turbulence.  This division is obviously
meant to be only a rough approximation of reality.

Numerical simulations of whole galaxies are still limited to space and
mass resolutions not much smaller than $\sim$1 kpc and $10^6$ \msun\
respectively (see, e.g., Weinberg, Hernquist \& Katz 2002; Steinmetz
\& Navarro 2002; Mathis et al. 2002; Lia, Portinari \& Carraro 2002;
Recchi et al. 2002; Pearce et al. 2001; Toft et al. 2002; Springel \&
Hernquist 2003; Tornatore et al. 2003; Governato et al. 2004).  They
can address effectively the processes dominant in the large-scale
range identified above, but the feedback processes acting on
intermediate and small scales are ``sub-grid'' physics and are treated
with simple heuristic models that require the introduction of free
parameters.

Current models of semi-analytic galaxy formation treat feedback at a
similar, phenomenological level (see, e.g., Cole et al. 2000;
Somerville, Primack \& Faber 2001; Diaferio et al. 2001; Poli et
al. 2001; Hatton et al. 2003); they typically connect the efficiency
of feedback to the circular velocity of the dark matter halo, with the
aid of free parameters.  Models of galaxy formation that include a
more detailed description of feedback have been presented, e.g., by
Silk (1997, 2001), Ferrara \& Tolstoy (2000), Efstathiou (2000), Tan
(2000), Lin \& Murray (2000), Hirashita, Burkert \& Takeuchi (2001),
Ferreras, Scannapieco \& Silk (2002) or Shu, Mo \& Mao (2003).  In
this framework (semi-)analytic work can give a very useful
contribution in selecting the physical processes that are most likely
to contribute to feedback.

The focus of this paper is on modeling the intermediate range of
scales defined above, where the physics of the ISM is in act.  The
standard picture of the ISM is that of a multi-phase medium in rough
pressure equilibrium; the reference model is that of McKee \& Ostriker
(1977), who considered a medium composed by cold, spherical clouds
with temperature and density \tc\circa 100 \K\ and \nc\circa 10 \cmt,
kept confined by a hot phase with \th\circa $10^6$ \K\ and \nh\circa
$10^{-3}$ \cmt.  A warm phase of $T_{\rm w}$\circa 10$^4$ \K\ and
$n_{\rm w}$\circa 10$^{-1}$ \cmt\ was produced at the interface.  This
vision is partially confirmed by multi-wavelength observations (see,
e.g., Heiles 2001), although reality appears more complex, suggesting
the presence of at least 5 different phases.

This picture is challenged by the results of many simulation programs,
aimed to the numerical modeling of the ISM (see, e.g., Mac Low et al.
1998; Ostriker, Gammie \& Stone 1999; Avila-Reese \& Vazquez-Semadeni
2001; Kritsuk \& Norman 2002; see Mac Low 2003 and Vazquez-Semadeni
2002 for reviews).  In this context the ISM is dominated by
compressible, supersonic, MHD turbulence.  These groups are still
struggling to tame the full complexity of the problem, so that these
simulations are not directly aimed to or easily usable by modeling of
galaxy formation.  For our purposes it is worth mentioning some
results.  The distributions of temperature and density of the
simulated gas particles show a wide range of values without any strong
multi-modality, but some broad peaks are anyway present. The
distribution of pressure shows a much more limited range of values.
Structures defined as overdensities are not static clouds but
transient features of an overall fractal distribution (which is
consistent with observations, see Chappell \& Scalo 2001) that do not
last more than a sound crossing time, unless they are gravitationally
bound.  Thus, the ``classical'' picture of the ISM is not validated,
but a model with multiple phases in rough pressure equilibrium can
still be used, though with care, as a useful first-order
approximation, able to catch some significant elements of the dynamics
of the ISM.

The motivation for the present work is to investigate the kind of
physical processes that arise in galaxy formation, in order to provide
a grid of solutions for the behaviour of feedback in a wide range of
realistic cases, to be used in simulations or semi-analytic models of
galaxy formation.  We restrict to a two-phase medium in pressure
equilibrium, composed by cold clouds embedded in a diffuse hot phase.
The dynamics of the ISM is at present assumed to depend only on its
``local'' properties, leaving thus out ``large scale'' events like
differential rotation, spiral arms, mergers, galactic winds and so on.
These events will be introduced once the global characteristics of the
galaxy are specified.

This paper is the first of a series aimed to modeling feedback in
galaxy formation.  It presents a minimal feedback model with its main
properties and results.  Preliminary results were presented by Monaco
(2002; 2003).  An upcoming paper will focus on the destruction of
collapsing, star-forming clouds (Monaco 2004, hereafter paper II).

The paper is organized as follows.  Section 2 describes the physical
ingredients of the model, Section 3 introduces the system of equations
used, Section 4 the main solutions.  Section 5 is devoted to a
discussion of the results, and Section 6 gives the conclusions.
Finally, three appendices give a list of frequently used symbols, a
determination of the time scales of coagulation of cold clouds and a
study of the fate of SBs in the \nh--\lmech\ plane.

\section{Feedback by steps}

Feedback is assumed to take place through a chain of processes:

(i) The densities and filling factors of the two phases are determined
by pressure equilibrium.

(ii) The cooled or infalled gas fragments into clouds with a given
mass spectrum; this is truncated at low masses (which are easily
destroyed) and at high masses (which continually collapse).

(iii) Collapse is triggered in clouds larger than the Jeans mass; we
use a criterion valid for non-spherical clouds.

(iv) Collapsing clouds are continually created by coagulation.

(v) Stars form in collapsing clouds.  Self-regulation of star
formation by HII regions destroys the clouds before most SNe explode.

(vi) SN remnants (hereafter SNRs) soon percolate into a SB, which
sweeps the ISM.  SBs heat the gas whenever they are in the adiabatic
stage, i.e. when the interior gas has not had time to cool, while they
collapse (and thus cool) the hot phase into a thin cold shell whenever
they get into the so-called snowplow stage.

(vii) SBs stop sweeping or collapsing the hot phase when they remain
pressure-confined or overtake the typical vertical scale-height of the
system (blow-out).

In the following we describe these steps in detail.  All distances are
given in \pc, masses in \msun, times in \yr, temperatures in \K, gas
densities in \cmt, average densities in \dens, surface densities in
\surf, energies in $10^{51}$ erg, mechanical luminosities in
$10^{38}$ erg s$^{-1}$, mass flows in \msunyr, energy flows in
$10^{51}$ erg/yr.  Pressures are divided by the Boltzmann constant
$k$ and given in \K\ \cmt.

\subsection{Pressure equilibrium}

Let's consider a volume $V$ filled with a two-phase medium, with
temperatures of hot and cold phases \th\ and \tc\ and densities \nh\
and \nc.  The volume is assumed to be large enough to contain many
star-forming clouds.  An external halo acts as a reservoir of gas,
which continually replenishes the cold component\footnote{ The halo is
assumed, for simplicity, to be completely decoupled from the hot
phase, although in realistic situations the two components will
interact.}.  Stars form from the cold gas.  The four components (cold
and hot phases, stars and the external halo) have masses \mcold,
\mhot, \mstar\ and \mhalo.  The total mass of the system is fixed to
\mtot.  The temperature of the cold gas is kept fixed to 100 \K,
i.e. roughly the position where the cooling function of the gas drops,
so that further cooling is inhibited unless the cloud collapses and
its density gets very high.  Let \muh\ and \muc\ be the mean molecular
weights of the two phases, \fh\ and \fc\ their filling factors
($\Fh+\Fc=1$), $\bar{\rho}_{\rm h}= \Mhot/V$ and $\bar{\rho}_{\rm c}=
\Mcold/V$ their average densities and $F_{\rm h}= \Mhot/(\Mcold +
\Mhot)$ the fraction of hot gas.  Pressure equilibrium implies:

\be \Nh\Th = \Nc\Tc\, . \label{eq:presseq} \ee

\noindent
From this we obtain:

\be \Fc = \frac{1}{1+\frac{F_{\rm h}}{1-F_{\rm h}}\frac{\Muc}{\Muh}
\frac{\Th}{\Tc}}\, , \label{eq:filcold} \ee

\noindent
and of course $\Fh=1-\Fc$, $\Nh=\bar{\rho}_{\rm h}/\Fh\Muh m_{\rm p}$ and
$\Nc=\bar{n}_{\rm c}/\Fc\Muc m_{\rm p}$ (where $m_{\rm p}$ is the proton mass).
Finally, the dependence of the \muh\ and \muc\ molecular weights on
metallicity is taken into account.

\subsection{Fragmentation of the cold phase}

It is assumed that the cold phase fragments into clouds with a given
mass spectrum.  As commented in the introduction, according to the
turbulent picture of the ISM the ``clouds'' (i.e. peaks of the fractal
density fields) are not stable entities but transient features of the
medium.  We will assume in the following that the self-gravitating
clouds are reasonably stable (in the sense that they are not
significantly reshuffled by turbulence) within one or two dynamical
times and that the continuous reshuffling of the density field does
not change the statistics of clouds.

The mass spectrum of the so-defined clouds is assumed to be a
power-law:

\be N\cl(m\cl)dm\cl = N_0 (m\cl/1\ {\rm M}_\odot)^{-\alpha\cl} dm\cl\, , 
\label{eq:cldistr} \ee

\noindent
where $N_0$ is a normalization constant (with dimensions pc$^{-3}$
M$_\odot^{-1}$), fixed by requiring $\bar{\rho_{\rm c}} = \int N\cl
m\cl dm\cl$ (see below), and \alfa\ is a free parameter.  This choice
is the natural outcome of many different processes, including
turbulence.  The parameter \alfa\ can be constrained both from theory
and observations of the ISM (see, e.g., Solomon et al. 1987), and
should vary between 1.5 and 2 (the latter considered as a reference
value), at least in self-regulated situations like the Milky Way.
Notice that in this way a significant amount of mass is located in
high-mass clouds.

To the clouds we associate a typical radius $a\cl$ defined simply as
$m\cl=4\pi a\cl^3\rho_{\rm c}/3$, or:

\be m\cl = 0.104\, \Muc\Nc a\cl^3\ {\rm M}_\odot\, . \label{eq:clradius} \ee

\noindent
This does not imply an assumption of sphericity of the clouds. 

The mass function of clouds is truncated both at low and high masses.
At the high mass end the mass function is truncated by gravitational
collapse, because clouds that form stars are quickly destroyed.  The
upper mass limit $m_{\rm u}$ will be computed in the next session.  At
low masses clouds are easily destroyed by a number of possible
processes, among which thermo- and photo-evaporation.  McKee \&
Ostriker (1977) set the lower limit to $a\cl = a_{\rm l} = 0.5$ pc.
For \muc\circa1.2 and \nc\circa10 \cmt\ this corresponds to $m_{\rm
l}$\circa0.1 \msun.  We set the lower mass limit to this value.  This
is surely a rough approximation, as $m_{\rm l}$ should be
self-consistently determined by the dynamics of the system, and is
unlikely to be a constant.  However, its actual value does not have a
strong impact on the results as long as $m_{\rm u}\gg m_{\rm l}$, a
condition that is verified by most solutions.  Nonetheless it is
important to set $m_{\rm l}$ to a non-vanishing value both to avoid
divergence in a few calculations (like the normalization of the mass
function for \alfa$\ge$2) and to avoid contributions from clouds that
most likely do not exist.

The normalization constant of the mass function is:

\be N_0\, (1\ {\rm M}_\odot)^{\alpha\cl} = \frac{\bar{\rho}_{\rm
c}}{f(m_{\rm u},m_{\rm l})}\, . \label{eq:clnorm} \ee

\noindent
Here the function $f(m_{\rm u},m_{\rm l})$ is equal to $(m_{\rm u}^{-\alpha\cl+2} -
m_{\rm l}^{-\alpha\cl+2}) /(-\alpha\cl+2)$ if $\alpha\cl\ne 2$, otherwise
$f(m_{\rm u},m_{\rm l})= \ln(m_{\rm u}/m_{\rm l})$.

\subsection{Critical mass for clouds}

Massive clouds are destroyed by gravitational collapse.  In absence of
magnetic fields and turbulence the threshold mass for collapse is
fixed by the Bonnor-Ebert criterion (Bonnor 1956; Ebert 1955), and
depends on an external pressure term $P_{\rm ext}$.  If the external
pressure is fixed to the thermal one, the criterion is equivalent to
the classical Jeans mass.  To generalize it to non-spherical clouds,
we follow Lombardi \& Bertin (2001), who find:

\begin{eqnarray} m_{\rm J} &\simeq& 1.18 \frac{c_{\rm s,c}^4}{\sqrt{G^3\Shape^3 P_{\rm ext}}}
\label{eq:lombert} \\ &\simeq& 20.3\, \Tc^{3/2} \Nc^{-1/2} \Muc^{-2} \Shape^{-3/2} \ {\rm M}_\odot\, .
\nonumber \end{eqnarray}

\noindent
Here $c_{\rm s,c}$ is the sound speed of the cold phase, the external
pressure is set to the thermal one and the parameter \shape\ is
defined by the authors as:

\begin{eqnarray}
\lefteqn{\Shape \equiv 12\pi \left(\frac{3}{4\pi}\right)^{1/3}\frac{V^{4/3}}{S^2}}
\label{eq:shape} \\
&& \times S^2 \left(\int_{\partial V} \left| \nabla_\xi u(s{\bf
x})\right|^{-1} dS 
\int_{\partial V} \left| \nabla_\xi u(s{\bf
x})\right| dS \right)^{-1}\, . \nonumber \end{eqnarray}

\noindent
In this equation the integrals are performed on the surface $\partial
V$ (of area $S$) of the volume $V$ of the cloud; the function $u$ is
the cloud density normalized to its maximum value
$u\equiv \rho/\rho_{\rm max}$ 
and $s^{-1}$ is a ``Jeans length'' defined as
$s=\sqrt{4\pi G \rho_{\rm max}/c_{\rm s,c}^2 }$; ${\bf x}$ is the space
coordinate and ${\bf \xi}=s {\bf x}$.  The parameter \shape\ is
dimensionless, scale invariant (i.e. does not change for similarity
transformations) and is always smaller than unity.  For a sphere
$\Shape=1$, and the Jeans (Bonnor-Ebert) criterion is recovered.  In
general, collapsing clouds will be non-spherical, and this will
correspond to an increase of the threshold mass $m_{\rm J}$.  We treat
\shape\ as a free parameter.  It can be considered as a product of two
terms, $\mu_1$ and $\mu_2$, given in the first and second lines of
Equation~\ref{eq:shape}.  Both terms are $\le 1$ and are unity for a
sphere; moreover, $\mu_2$ is unity when gravity is negligible.  So, a
rough estimate can be obtained by considering \shape\circa$\mu_1$.

This quantity is easily computed in the simple case of a rotational
ellipsoid with semi-axes $a_1$ and $a_2$ (with the third semi-axis
$a_3=a_2$).  If $r=a_2/a_1$ is the axial ratio, we find:

\be \Shape \simeq \frac{1}{g(r)^2r^{4/3}}\, , \label{eq:ellips} \ee

\noindent
where $g(r)=1/2+\arcsin\sqrt{1-r^2}/2r\sqrt{1-r^2}$ if $r<1$ and
$g(r)=1/2+\log [(r+\sqrt{r^2-1})/(r-\sqrt{r^2-1})]/4r\sqrt{r^2-1}$ if
$r>1$.  In this case \shape\ takes values \circa0.5 for axial ratios
of order 1:5 (in both senses), while it gets to \circa0.2 for axial
ratios 1:10.  As this is likely to be an overestimate of the actual
value, we consider 0.2 as a reference value for this parameter.

Magnetic fields and turbulence could in principle invalidate the
Bonnor-Ebert criterion by providing non-thermal support to the cloud.
Recent simulations (see, e.g., Mac Low 2003) have shown that
turbulence cannot inhibit the collapse of critical clouds; the Jeans
criterion remains valid provided that the quadratic sum of kinetic and
sound speeds is used in place of the sound speed itself.  For a
typical turbulent speed of several \kms, the Jeans mass would
correspond to that relative to a temperature \tc\ of several 10$^3$
\K.  The effect of turbulent motions can thus be roughly implemented
by assuming a very small value for \shape, of order 0.01.  Magnetic
fields can halt the global collapse of the cloud but not its
fragmentation into stars, so their effect on the critical mass for
collapse is negligible.

Finally, in cases like the sweeping of a spiral arm or during a merger
the Jeans criterion can be changed by explicitly introducing a $P_{\rm
ext}$ term.  This will correspond to a sudden decrease of the Jeans
mass, and then to a burst of star formation.

\subsection{Coagulation of cold clouds}

Clouds larger than the Jeans mass are continually created by kinetic
aggregation (coagulation) of smaller clouds.  This is described with
the aid of the Smoluchowski equation (von Smoluchowski 1916).  In this we
follow the approach of Cavaliere, Colafrancesco \& Menci (1991; 1992;
see also Menci et al. 2002), who used this formalism to describe the
kinetic aggregation of dark-matter halos.

The details of the calculations are reported in Appendix B.
In brief, the coagulation of clouds is driven by a kernel:

\be K = \bar{\rho}_{\rm c} \left \langle\left \langle\Sigma_{\rm coag}
v_{\rm ap} \right\rangle_{\rm v} \right\rangle_{\rm m}\, . \label{eq:kernel} \ee

\noindent
Here $\Sigma_{\rm coag}$ is the cross-section for interaction and
$v_{\rm ap}$ is the approach velocity, while the two averages are done
over velocity and mass.  Notably, it is assumed that clouds, although
transient, are stable for one crossing time $a\cl v_{\rm ap}$; this is
reasonable as $v_{\rm ap}$ is typically larger than the sound speed of
the cold phase.  Following Saslaw (1985) the cross-section for the
coagulation of two clouds (denoted by 1 and 2) is:

\be \Sigma_{\rm coag} = \pi (a_1+a_2)^2\left(1+2G\frac{(m_1+m_2)}
{a_1+a_2} \frac{1}{v_{\rm ap}^2}\right)\, . \label{eq:crosssect} \ee

\noindent
The first term corresponds to geometric interactions, the second to
resonant ones; this last term is effective when the approach velocity
is not much larger than the internal velocity dispersion of the
clouds.  In most cases considered here the geometrical term results
dominant, so we will neglect resonant interactions in the following.
Notice that this cross-section is valid for spherical clouds; we do
not consider the effect of asphericity here, as it would be a
further-order correction with respect to that of the Jeans mass
introduced above.

It is shown in Appendix B that the time scale for coagulation is:

\be t_{\rm coag} = \left(\frac{4\pi}{3}\right)^{2/3} \frac{1}{\pi}
\bar{\rho}_{\rm c}^{-1/3} \frac{\rho_{\rm c}}{\bar{\rho}_{\rm c}}^{2/3}
\frac{m_{\rm J}^{1/3}}{\langle v_{\rm ap} \rangle
} \label{eq:coag1} \ee

\noindent
The typical mass scale of the mass function, identified with the upper
cutoff, grows like $(1+t/3t_{\rm coag})^3$.  For a Maxwellian
distribution of velocities with 1D dispersion \sigv\ we have $\langle
v_{\rm ap} \rangle = 1.30 \sigma_{\rm v}$.

The time at disposal for accretion is the time necessary to a
Jeans-mass cloud to be destroyed.  This will be related to the
dynamical time:

\be \Tff = \sqrt{\frac{3\pi}{32G\rho_{\rm c}}} \simeq 5.15 \times 10^7\,
(\Muc\Nc)^{-1/2} \ {\rm yr}\, . \label{eq:freefall} \ee

\noindent
As star formation is triggered roughly after \tff, and early feedback
from young stars destroys the cloud in a comparable time (see below),
we conservatively allow aggregation to go on for two dynamical times.
Thus, the upper mass cutoff is set to:

\be m_{\rm u} = m_{\rm J} 
\left(1+\frac{2\Tff}{3t_{\rm coag}}\right)^3
\, . \label{eq:upper} \ee

The mass of the typical collapsing cloud is then:

\be \Mcc = \frac{\int_{m_{\rm J}}^{m_{\rm u}} m\cl N\cl(m\cl) dm\cl}
{\int_{m_{\rm J}}^{m_{\rm u}} N\cl(m\cl) dm\cl}\, , \label{eq:mcc} \ee

\noindent
and the fraction of cold gas presently available for star formation
is:

\be \Fcoll = \frac {\int_{m_{\rm J}}^{m_{\rm u}} m\cl N\cl(m\cl) dm\cl} 
{\bar{\rho}_{\rm c}}\, . \label{eq:fcoll} \ee

\noindent
The total number of collapsing clouds is:

\be \Ncc =  \Fcoll \frac{\Mcold}{\Mcc} \label{eq:ncc} \ee


Coagulation is a physically motivated and reasonable mechanism to
explain the growth of cold clouds, but it has never been validated (to
the best of our knowledge) by simulations that include MHD turbulence.
Besides, it has been proposed that giant molecular clouds form in the
converging flows caused by the sweeping of spiral arms
(Ballesteros-Paredes, Vazquez-Semadeni \& Scalo 1999), a process that
cannot be introduced without a proper modeling of the disc.

A consequence of the assumptions done is that cooling alone is not
going to produce clouds larger than the Jeans mass; they are produced
only by coagulation of smaller clouds.  This is contrary to the naive
expectancy of a mass function of clouds which is truncated {\it below}
by the Jeans mass in case of a cooling flow, as only fluctuations
larger than the Jeans mass can grow.  This is not what is observed in
the case of thermal instability in turbulent media, where (without
thermal conduction and UV heating) structures of all masses are
observed down to the resolution limit (see, e.g, Kritsuk \& Norman
2002).  On the other hand, it is possible that giant clouds, much
larger than the Jeans mass, form in the cooling flows that take place
at the centres of cosmological halos.  This is neglected here, but can
be modeled by introducing a further mass scale in the mass function.

\subsection{Star formation and early feedback}

Collapsing clouds can reach high enough densities to trigger the
formation of H$_2$ and further cool to $\sim$10 \K.  After one
dynamical time (Equation~\ref{eq:freefall}) star formation starts
inside the ``molecular'' cloud.  An important point is that early
feedback from massive stars can destroy the collapsed cloud before the
bulk of type II SNe has exploded.  It has been shown (Franco, Shore \&
Tenorio-Tagle 1994; Williams \& McKee 1997; Matzner 2002) that HII
regions are a source of turbulence, and their energy input is
sufficient to destroy the star-forming clouds, pre-heating them at
$\ga$$10^4$ \K.  A similar role is played by stellar winds, that are
typically trapped inside HII regions (McKee, van Buren \& Lazareff
1984).  Matzner (2000) computed the amount of turbulence driven into
the star-forming cloud by expanding HII regions.  Under the assumption
that the rate of injection of turbulence equates the decay rate
estimated from N-body simulations, he predicted that the cloud would
be destroyed in \circa$2\times 10^7$ \yr, i.e. about one dynamical
time of the uncollapsed cloud (Equation~\ref{eq:freefall} with \nc\circa10
\cmt), with a resulting efficiency of star formation \fstar\ (i.e. the
fraction of the cloud that goes into formed stars) of \circa5-10 per
cent.  This is in rough agreement with both observations of molecular
clouds and estimates from globular clusters (\fstar\circa1--10 per
cent; see, e.g., Elmegreen 2000, 2002).

The ability of the energy from SNe to emerge from the destroyed cloud,
possibly the most delicate step in the whole chain of feedback events,
is addressed in paper II; here we give only a very short summary of
the results.  When SNe start to explode the cloud is already in the
process of being destroyed, so that a significant fraction of mass is
in a warm, diffuse phase.  SNRs propagating in this dense environment
soon radiate their thermal energy (see next section for more details).
In this case, the mass internal to the blast collapses into a thin,
dense shell that fragments as soon as the blast is confined by kinetic
pressure.  So, the net effect of the first SNe is that of collapsing
again the diffuse material heated up by the HII regions.  After a few
SNe, most gas is re-collapsed into cold clouds with a low filling
factor, while the diffuse component has such a low density that SNRs
emerge from the cloud before cooling.  From this point all the energy
from SNe is used to drive the SB.  In case many tens of SNe explode in
a single cloud, most energy (90-95 per cent) is used to drive the SB,
while for very small clouds, where only a few SNe explode, the first
SN is able to destroy the cloud, losing most of its energy in the
process, while the other SNe (if any) will pump energy into the ISM
with a likely high efficiency.  Eventually, only \circa10 per cent of
the initial cloud is found in diffuse, hot gas with temperature of
order 10$^6$ \K; lower values are expected if the cloud is
particularly dense.

For this version of the feedback model we decide to give a minimal,
heuristic description of this process, in order to keep the model
simpler.  Each SN releases $10^{51}E_{51}\ erg$ in the
ISM\footnote{Observations suggest values of \esn\ in the range 1 to
10.}.  We assume that all the energy is available for driving the SB;
in case of very small collapsing clouds a lower effective value of
\esn\ will be plausible.  We assume that a fraction \fevap\ of the
cloud is evaporated to a temperature \tevap, while the rest (amounting
to a fraction $1-\Fstar-\Fevap$) is re-collapsed into cold clouds.  Of
course \fevap+\fstar$\le$1.  We use as reference values \esn=1,
\fevap=0.1\ and \tevap=$10^6\ K$, with the warning that in case of
very dense clouds \fevap\ will likely be lower (see paper II).

Finally, the contribution of a single collapsing cloud to the global
star formation rate is:

\be \dot{m}_{\rm sf} = \Fstar \frac{\Mcc}{\Tff}\, . \label{eq:clsfr} \ee

\begin{table*}
\begin{center}
\begin{tabular}{l|lcl}
\hline
&&&Adiabatic stage \\
\hline
Radius & $\Rsbad(t)$ &=& $81.3\ (L_{38}/\Muh\Nh)^{1/5}
t_6^{3/5}$ pc \\
Shock speed & $v_{\rm sb}^{\rm (ad)}(t)$ &=& $47.7\ (L_{38}/\Muh\Nh)^{1/5} 
t_6^{-2/5}$  km s$^{-1}$\\
Average temp. & $\bar{T}_{\rm sb}^{\rm (ad)}(t)$ &=& $1.79\times10^5\
L_{38}^{2/5} \Muh^{3/5}\Nh^{-2/5}t_6^{-4/5}$ K \\
Post-shock temp. & $T_{\rm sb}^{\rm (ad)}(t)$ &=& $5.4\times10^4\
L_{38}^{2/5} \Muh^{3/5}\Nh^{-2/5}t_6^{-4/5}$ K \\
Post-shock press. &$P_{\rm sb}^{\rm (ad)}(t)/k$ &=&
$2.16\times10^5\ L_{38}^{2/5} (\Muh\Nh)^{3/5} t_6^{-4/5}$ K cm$^{-3}$ \\
Cooling time & $t_{\rm cool}^{\rm (ad)}(t)$ &=& $255\ L_{38}^{3/5} \Muh^{9/10}
\Nh^{-8/5} \zeta_{\rm m}^{-1} t_6^{-6/5}$ yr\\
Swept mass & $M_{\rm sw}(t)$ &=& $5.53\times10^4\ L_{38}^{3/5}
(\Muh\Nh)^{2/5} t_6^{9/5}\ {\rm M}_\odot$\\
Internal mass & $M_{\rm int}(t)$ &=& $M_{\rm sw}$\\
\hline
&&&PDS stage\\
\hline
Radius & $\Rsbpds(t)$ &=& $70.2\ (L_{38}/\Muh\Nh)^{1/5}
t_6^{3/5}$ pc \\
Shock speed & $v_{\rm sb}^{\rm (pds)}(t)$ &=& $41.2\ (L_{38}/\Muh\Nh)^{1/5} 
t_6^{-2/5}$  km s$^{-1}$\\
Post-shock press. &$P_{\rm sb}^{\rm (ad)}(t)/k$ &=&
$1.60\times10^5\ L_{38}^{2/5} (\Muh\Nh)^{3/5} t_6^{-4/5}$ K cm$^{-3}$ \\
Swept mass & $M_{\rm sw}(t)$ &=& $3.56\times10^4\ L_{38}^{3/5}
(\Muh\Nh)^{2/5} t_6^{9/5}\ {\rm M}_\odot$\\
Internal mass & $M_{\rm int}(t)$ &=& $M_{\rm sw}\
(1-(t/\Tpds)^{-3.2})$\\
\hline
\end{tabular}
\label{table:sb}
\caption{Main properties of SBs.}
\end{center}\end{table*}

\subsection{Super-bubbles}

SNRs associated to massive stars in a star-forming cloud will soon
percolate into a single hot bubble.  As a consequence, all the SNe
exploding in a cloud will drive a single SB into the ISM (see, e.g.,
Mac Low \& McCray 1988).

Stars are formed with a given Initial Mass Function (hereafter IMF)
that must be specified.  For the model the only information needed is
the mass of stars formed for each supernova, \mstsn.  We associate one
SN to each $>8\ {\rm M}_\odot$ star; if the (differential) IMF has a slope
$-(\alpha_{\rm imf}+1)$ and the lifetime of a star goes like its mass
raised to $-\alpha_{\rm life}$, the rate of SN explosion goes like
$t^{(\alpha_{\rm imf}-\alpha_{\rm life})/\alpha_{\rm life}}$.  For
standard choices of $\alpha_{\rm imf}=1.35$ and $\alpha_{\rm life}\sim
2.5 - 3$ the exponent takes a value of $\sim -0.5$.  In other words,
the rate of SN explosion depends weakly on time, and is approximated
as constant.  Denoting by \tlife\ the difference between the lifetime
of an $8\ {\rm M}_\odot$ star and that of the largest star, the number of SNe
that explode in a collapsing cloud and the resulting rate are:

\be \Nsn = \Fstar \frac{\Mcc}{\Mstsn}\, , \label{eq:nsn} \ee

\be \Rsn = \Fstar \frac{\Mcc}{\Tlife\Mstsn}\, . \label{eq:rsn} \ee

\noindent
The mechanical luminosity of the SB is then $L_{\rm mech} = L_{38} \times
10^{38}\ {\rm erg\ s}^{-1}$, where:

\be L_{38} = \frac{10^{13}\Rsn E_{51}}{1\ {\rm yr}}\, . \label{eq:l38}
\ee

In presence of a two-phase medium the SB expands into the more
diffuse, more pervasive hot phase; cold clouds will pierce the
blast, but this will promptly reform after the cloud has been
overtaken (McKee \& Ostriker 1977; Ostriker \& McKee 1988; Mac Low \&
McCray 1988).  

The evolution of the SB is described following the model of Weaver et
al. (1977; see also Ostriker \& McKee 1988).  In the beginning the SB
is adiabatic, because the shocked ISM has not had time to cool.  In
this case:

\be \Rsbad(t) = 81.3 \left( \frac{L_{38}}{\Muh\Nh}\right)^{1/5}
t_6^{3/5}\ {\rm pc}\, , \label{eq:rsbad} \ee

\noindent
where $t_6 = t/10^6$ yr.  Table 1 reports the main properties of the
SB expanding in the hot phase.  

In the adiabatic stage the hot phase is shock-heated by the blast.  Of
the initial energy of the SN, 73.7 per cent is thermal and 26.3 per
cent is kinetic.  This stage ends when the post-shock mass elements
cool, i.e. when $t_{\rm cool}^{\rm (ad)}(t)=t$.  The cooling time is
computed as:

\be t_{\rm cool}=3kT/\Nh \Lambda(T) \label{eq:tcool} \ee  

\noindent
and is evaluated at $T_{\rm sb}^{\rm (ad)}$ (given in Table 1) and
$4\Nh$ (due to the shock jump condition).  For the cooling function we
use the approximation proposed, e.g., by Cioffi, McKee \& Bertschinger
(1988):

\be \Lambda = 1.6\times 10^{-19}\, \zeta_{\rm m} \Th^{-1/2}\, , \label{eq:coolfunc}
\ee

\noindent
where $\zeta_{\rm m}\equiv \Zhot/Z_\odot$ is the metallicity of the hot
gas in solar units.  This formula is relatively accurate in the range
$10^5 \le \Th \le 10^{6.5}$.  A more realistic cooling function would
be desirable, but would make analytic estimates unfeasible.
The time of shell formation is then:

\be \Tpds = 2.33 \times 10^4 L_{38}^{3/11} \Muh^{9/22} \Nh^{-8/11}
\zeta_{\rm m}^{-5/11}\ {\rm yr} \label{eq:tpds} \ee

\noindent
We call \rpds\ the radius of the SB at \tpds.  After this moment the
swept mass collapses into a thin cold shell.  This shell acts like a
snowplow, making the swept ISM collapse into it.  For simplicity the
cold clouds are assumed as before to pierce the shell without any
effect.  Some of the hot gas will anyway remain inside the bubble,
pushing the snowplow with its pressure; this stage is called Pressure
Driven Snowplow (PDS).  We use the solution of Weaver et al. (1977;
see also Castor, McCray \& Weaver 1975) that includes thermal
conduction at the interface between the gas and the cold shell, a
mechanisms that releases more hot gas from the shell into the
interior.  They obtain:

\be \Rsbpds(t) = 70.2 \left( \frac{L_{38}}{\Muh\Nh}\right)^{1/5}
t_6^{3/5}\ {\rm pc}\, , \label{eq:rsbpds} \ee

\noindent
Notice that the time dependence is the same as above, due to the
presence of an increasing amount of hot interior gas.  This gas is
however negligible with respect to the swept mass, and is so diluted
that further cooling is inhibited.  A more standard choice for the
evolution of the SB in the PDS stage would be (see, e.g., Koo \& McKee
1992) $\Rsbpds\propto t^{4/7}$; the exponent decreases only by 5 per
cent with respect to Weaver et al. (1977).  Table 1 reports the main
characteristics of the SBs in this stage.

To ease numerical integration we interpolate between the adiabatic and
PDS stages assuming that after \tpds\ the blast radius evolves like
$R_{\rm sb}\propto t^{0.2}$ and the velocity like $v_{\rm sb}\propto
t^{-2}$ until the PDS solutions are met.

In the PDS stage, the amount of ISM swept by the SB that is collapsed
into the shell is estimated as the fraction of the internal material
for which (in the adiabatic solution) $t_{\rm cool}(r;t)<t$.  Assuming
a power-law profile for density and temperature of the gas just inside
the adiabatic blast, in the pressure-gradient approximation of
Ostriker \& McKee (1988) we obtain $T\propto (r/R_{\rm sb})^{0.5}$ and
$\rho\propto (r/R_{\rm sb})^9$.  From these relations we obtain that
the internal mass (not yet collapsed into the shell) is related to the
swept mass as:

\be M_{\rm int}= M_{\rm sw} (t/\Tpds)^{-3.2}\, .\label{eq:mint}
\ee

\noindent
This is valid of course for $t>\Tpds$.  For simplicity we assume that
the thermal energy of the SB is lost at the same rate:

\be \Eth = 0.737\, \Rsn t \left(1-(t/\Tpds)^{-3.2}\right)\, ,
\label{eq:eth} \ee

\noindent
while the kinetic energy is kept at $\Ekin=0.263\Rsn t$.

The explosion of the last SN marks the exhaustion of energy injection
into the SB, so the evolution after this event should follow that of a
SNR.  We observe that SBs are stopped by thermal pressure or by
blow-out (see below) before exhaustion in virtually all cases, so an
accurate modeling of this stage is immaterial.  In any case, we assume
that after the last SN has exploded the blast always evolves like the
adiabatic Sedov solution for a SNR, $R_{\rm sb}\propto t^{2/5}$.


A note of caution is necessary on the application of these solutions
for the evolution of the SB.  They are valid if the hot phase is
uniform, the cold phase negligible, and if the mass of the ``wind''
that drives the SB is negligible with respect to the swept mass.  This
last condition is violated in most actual cases as soon as a
significant fraction of the collapsing cloud is evaporated.  On the
other hand, the other conditions are also violated: the ISM is
structured, and events like thermo-evaporation of clouds, dragging of
clouds by the internal gas, turbulent and magnetic pressure and cosmic
rays are likely to influence significantly the dynamics of the SB.
Some of these effects can be modeled analytically (see Ostriker \&
McKee 1988), but at the cost of more uncertainties and dangerous
assumptions.  We decide to rely on the simple solutions given above,
with the {\it caveat} that all numbers must be considered as useful
order-of-magnitude estimates.

Nonetheless, we have tried to include thermo-evaporation of the cold
phase by the expanding SB, so as to quantify the mass flow implied.
Following the approach of McKee \& Ostriker (1977) and the
generalization of Ostriker \& McKee (1988) to SBs, we find that the
thermo-evaporated ISM is generally much less massive than the
evaporated gas from the star-forming cloud, so the inclusion of
thermo-evaporation, while introducing further uncertainties, does not
affect strongly the results presented here.  We will neglect
thermo-evaporation in the following.

\subsection{The fate of SBs}

An SB can end in two ways: (i) being confined by external pressure,
(ii) blowing out of the system.

Case (i) takes place at time \tconf\ when the shock speed is equal to
the external, thermal one:

\be v_{\rm sb}(\Tconf) = c_{\rm s,h} = 91.2 (T_{\rm h,6}/\Muh)^{1/2},
\label{eq:conf} \ee

\noindent
where $T_{\rm h,6}=\Th/10^6\ K$.  As the blast propagates into the
low-density hot phase, kinetic pressure is always negligible.  The
time and radius at which confinement takes place are given in Table C1
of Appendix C.  After confinement, the blast (in the adiabatic stage)
dissolves in the hot phase or the shell (in the PDS stage) fragments
because of Raileigh-Taylor instabilities.  This allows the hot phase
to mix with the interior hot gas.  However, as long as $\Tconf<\Tlife$
many SNe explode after confinement.  This will correspond to the
creation of secondary bubbles; the medium in which they expand will
depend on the velocity with which the interior gas mixes with the
external one.  In the adiabatic confinement case, it is easy to see
that secondary bubbles will be confined in the adiabatic stage as
well, so all the energy from SNe will be released to the hot phase; in
this case feedback is mostly efficient.  In case of confinement in the
PDS stage, the situation is more complicated.  If interior and
external hot gas mix very quickly, the secondary bubbles will expand
in the same medium and will then create secondary shells, but if mixing
is slow the energy of the remaining SNe will be pumped efficiently
into the hot, rarefied internal gas.  To address this case we assume
that the energy of the SNe exploding after confinement is released to
the hot phase either entirely, \fpds=1, or by a fraction
\fpds=$0.737(1 -(\Tconf/\Tpds)^{-3.2}) +0.263$, that takes into
account that thermal energy is dissipated according to
Equation~\ref{eq:eth}.  The two cases should bracket the true
solution.

Case (ii), the blow-out of the SB, takes place when the SB overtakes
the vertical scale-height \heff\ of the system, defined as (Mac Low \&
McCray 1988; Koo \& McKee 1992):

\be \Heff \equiv \frac{1}{\rho_0}\int_0^\infty\rho_{\rm h}(z)dz\, ,
\label{eq:heff}\ee

\noindent
where $z$ is the vertical direction (that for which \heff\ is minimal)
and $\rho_0=\rho_{\rm h}(z=0)$.  The blow-out condition is obviously
$R_{\rm sb}=\Heff$, and the blow-out time is reported in Table C1.
This condition is true if all SBs are centred at $z=0$.  Coagulation
naturally leads to mass segregation, and this is in line with the
observational evidence that molecular clouds show a smaller vertical
scale-length then HI.  However, typical blowing-out SBs will be away
from the mid-plane and will blow out only from one side, with a result
that is not vastly different from a bi-polar blow-out.  Moreover,
off-plane SBs will blow out more easily, and intermediate
configurations with mid-plane SBs being pressure-confined and external
SBs blowing out will be possible.  In order to keep the model as
simple as possible we consider only mid-plane SBs; the \heff\ vertical
scale-length will then be understood as the difference between the HI
and H$_2$ scale-lengths

The SB does not stop immediately after blow-out, as the rarefaction
wave that follows blow-out must have time to reach the blast traveling
in the horizontal direction.  We then allow for a sound crossing time
before stopping the SB.  If $\alpha^2$ is the mean effective Mach
number of the blast (the square ratio between the blast speed and the
average internal sound speed) and $R_{\rm sb}(t)\propto t^\eta$, the
sound crossing time of an adiabatic bubble is:

\be t_{\rm cross} = \frac{\alpha}{\eta}\Tbo\, ; \label{eq:cross} \ee

\noindent
For adiabatic and PDS blasts, $\alpha^2=1.61$ and 1.18 (Ostriker \&
McKee 1988; Weaver et al. 1977).  The final radius \rbos\ and time
\tbos\ ($\equiv\Tbo+t_{\rm cross}$) at blow-out are reported in table
C1.  Between \tbo\ and \tbos\ the SB can get into the PDS stage (if it
hasn't yet) or be confined by pressure.

The final time \tfin\ and radius \rfin\ are defined respectively as
the smallest between \tbos\ and \tconf, and between \rbos\ and \rconf.
Appendix C reports a study of the final state of SBs (confinement or
blow-out in adiabatic or PDS stage) in the plane defined by the two
variables \nh\ and \lmech.

\begin{figure}
\centerline{
\includegraphics[width=8.4cm]{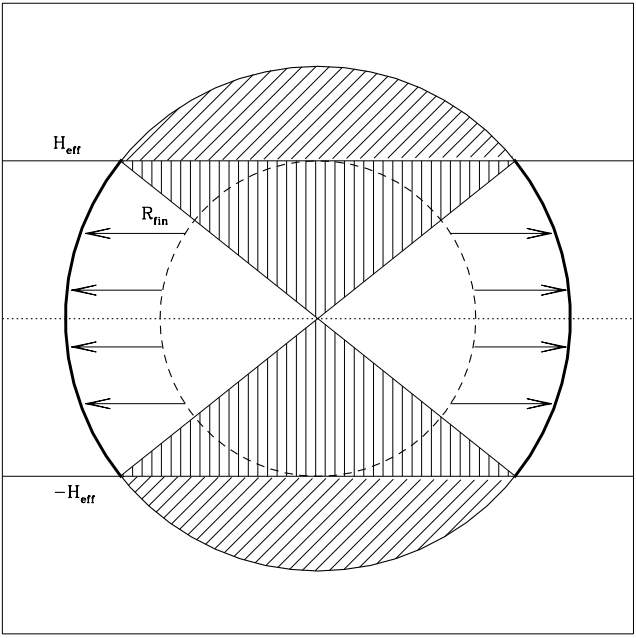}
}
\caption{Geometrical model for blow-out.  The SB starts blowing out
when its radius is equal to \heff, but continues to expand for one
sound crossing time, finally reaching the radius \rfin.  The two polar
cups (diagonal-shaded regions), defined by the intersection of the
final SB and the two planes at \heff, are assumed to be devoid of
matter.  All the matter present in the double cone (vertical-shaded
regions) with an aperture equal to that of the polar cups receives a
radial momentum that allows it to blow-out into the halo.}
\label{fig:blowout}
\end{figure}

At blow-out part of the hot interior gas of the SB escapes to the
halo.  To compute the fraction of hot blown-out gas we adopt the
following simple geometrical model (Fig.~\ref{fig:blowout}).  The ISM
swept by a SB of radius \rfin\ blows out from the two polar cups
defined by the intersection of the sphere and the two horizontal
planes at distance \heff\ from the centre.  The swept gas receives
momentum from the blast in the radial direction, so the blowing-out
gas is that contained in a double cone with the opening angle $\theta$
of the polar cups; we have $\cos\theta = \Heff/\Rfin$.  Neglecting the
ISM contained in the polar cups (that are outside the volume $V$), the
fraction of swept ISM that is blown out is:

\be \Fbo = \left\{ 
\begin{array}{lll} \frac{1}{2}[\Heff/\Rfin - (\Heff/\Rfin)^3] & {\rm if}
& \Rfin > \Heff \\ 0 & {\rm if} & \Rfin < \Heff\, . \end{array} \right.
\label{eq:fbo} \ee

\noindent
This is valid both for adiabatic and PDS blow-out.  Consistently, the
absence of ISM in the two polar cups is considered when computing the
swept mass.  With this simple model, that contains no free parameters,
the fraction \fbo\ ranges from 0 to \circa0.2; this is roughly
consistent with Mac Low \& McCray (1988) and Mac Low, McCray \& Norman
(1989), who report that most of the internal hot gas remains in the
disc.  It is anyway interesting to allow for higher \fbo\ values; this
is done by forcing the maximum of Equation~\ref{eq:fbo} to be \fbom,
which is taken as a free parameter.

A note on definitions: in a cosmological context the term blow-out is
used for the gas that is expelled from a galactic halo; we use it for
the expulsion of gas from the ``galaxy'', i.e. from the region where
stars and ISM are present, but our blown-out gas is destined by
construction to remain in the halo.  As already mentioned above, this
over-simplification is introduced to avoid modeling of the external
halo.  As we know temperature, density and escape velocity of the
blown-out gas, modeling of galactic winds is readily feasible once the
global properties of the hosting dark-matter halo are specified.

The energy of the SNe exploding after \tbos\ is assumed to be funneled
out into the halo.  Nonetheless, the restored mass, responsible for
chemical enrichment, is for simplicity blown out with the same
efficiency (\fbo) as the rest of the mass, leading to a possible
underestimate of the metals ejected into the halo.  We will see in the
following that the ejection of metals in the halo is anyway rather
efficient even with this assumption (see also de Young \& Gallagher
1990; Ferrara \& Tolstoj 2000).  We leave a refinement on this point
to further work.

The porosity \poro\ of the SBs is defined as the fraction of volume
occupied by expanding blasts.  This is a very important quantity, as
when it is unity it indicates that the blasts percolate the volume and
create a super-SB.  The computation of \poro\ depends on the time at
which SBs stop to expand into the ISM.  In case of blow-out the blast
halts at \tbos\ and the energy of further SNe is funneled out of the
volume $V$ into the halo.  On the contrary, in case of confinement
secondary blasts form after \tfin, whose energy is still injected into
the ISM.  To take this into account we compute \poro\ as:

\be \Poro = \frac{\Ncc}{\Tff V} \frac{4}{3}\pi \int_0^{t_{\rm poro}} R_{\rm sb}^3(t) dt\, .
\label{eq:poro} \ee

\noindent
where $t_{\rm poro}=\Tfin$ in case of blow-out, or \tlife\ in case of
confinement; in the latter case $R_{\rm sb}$ is kept constant to
\rfin\ after \tfin.  Regarding the adiabatic blow-out regime, it is
clear that if SBs remain identifiable after $t_{\rm poro}$, then a
value of \poro\ inferred from observations will be higher than that
given by Equation~\ref{eq:poro}.  However, recognizable bubbles do not
play the same dynamical role as expanding blasts.

\section{The system of equations}

\subsection{Mass flows}

\begin{figure}
\centerline{
\includegraphics[width=8.4cm]{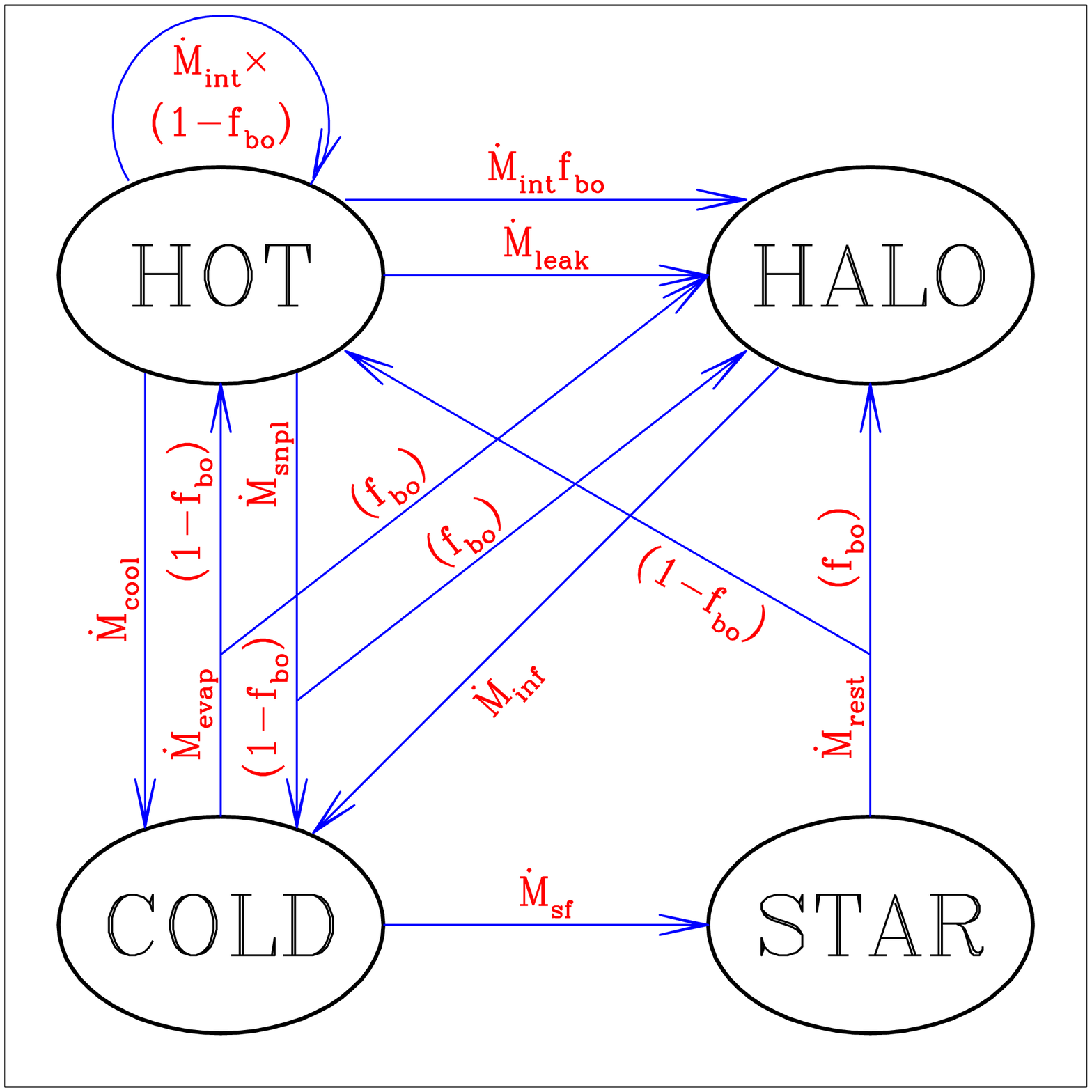}
}
\caption{Mass flows between the four components described in the
model.  Arrows denote the flows connected to infall (\dminf), star
formation (\dmsfr), restoration (\dmrest), cooling (\dmcool),
evaporation (\dmev), snowplows (\dmsnpl), leak-out (\dmleak) and the
rate at which the hot phase is engulfed by SBs (\dmint).  Blow-out
takes mass by a fraction \fbo\ from the internal, evaporation,
snowplow and restoration mass flows.}
\label{fig:fluxes}
\end{figure}

Fig.~\ref{fig:fluxes} shows all the mass flows between the four
components that are taken into account in this model.  

Cold gas is continually infalling from the halo.  This is modeled
simply as follows:

\be \Dminf = \frac{\Mhalo}{\Tinf}\, , \label{eq:dminf} \ee

\noindent
where \tinf\ is a parameter of the system.

The hot phase cools at the rate \tcool; if it were completely
homogeneous it would remain globally hot.  In realistic cases the hot
phase will show a rather broad range of densities and local cooling
times, so that a fraction of the gas will be able to cool to low
temperature.  A modeling of this fraction would require detailed
knowledge of the density distribution of the hot phase; we prefer to
leave it as a free parameter, \fcool, with 0.1 as a tentative
reference value.  The cooling mass flux is then:

\be \Dmcool = \Fcool \frac{\Mhot}{\Tcool}\, , \label{eq:dmcool} \ee

\noindent
where $\Tcool(\Th)$ is computed from Equation~\ref{eq:tcool} and with
the simple cooling function given by Equation~\ref{eq:coolfunc}.

While the cold phase is easily confined by a modest gravitational
well, the hot phase is generally able to leak out of the volume $V$ to
the halo.  The time scale connected to this leak-out is:

\be \Tleak = \sqrt{3/d} \frac{\Heff}{c_{\rm s,h}}\, ,
\label{eq:tleak} \ee

\noindent
where $c_{\rm s,h}$ is the sound speed of the hot phase and $d$ is
unity if leak-out is on one preferential direction, 3 if it is
spherically symmetric.  In the following we will consider for
simplicity leak-out in one direction; we have verified that the
results are not sensitive to the value of $d$.  The mass loss rate is
then:

\be \Dmleak = \frac{\Mhot}{\Tleak}\, . \label{eq:dmleak} \ee

\noindent
This term should be revised if some external hot halo gas hampers
leak-out.

Cool gas transforms into stars at the rate:

\be \Dmsfr = \Fstar \Fcoll \frac{\Mcold}{\Tff}\, . \label{eq:dmsfr} \ee

\noindent
This is easily obtained by multiplying the contribution of a single
cloud (Equation~\ref{eq:clsfr}) by the total number of collapsing
clouds \ncc\ (Equation~\ref{eq:ncc}).  It can be written also as
$\Dmsfr=\Mcold/\Tsf$, where

\be \Tsf = \Tff/\Fstar \Fcoll \, . \label{eq:tsf} \ee

A fraction \frest\ is instantaneously restored to the hot phase:

\be \Dmrest = \Frest \Dmsfr\, . \label{eq:dmrest} \ee

\noindent
This flux is responsible for chemical enrichment; we notice that this
equation implies instantaneous recycling.

The rate at which the mass of collapsing clouds is evaporated back to
the hot phase is:

\be \Dmev = \Fevap \Fcoll \frac{\Mcold}{\Tff}\, .
\label{eq:dmev} \ee

\noindent
It follows that $\Dmev=\Fevap\Dmsfr/\Fstar$.

At the final time \tfin\ each SB has swept a mass $M_{\rm sw}(\Tfin)$,
of which a part $M_{\rm int}(\Tfin)$ (see Table 1 and
Equation~\ref{eq:mint}) is in hot internal gas and the rest is in the
snowplow.  The rate at which the hot phase becomes internal mass of a
SB is:

\be \Dmint = \Ncc \frac{M_{\rm int}(\Tfin)}{\Tff}\, , \label{eq:dmint}
\ee

\noindent
while the rate at which it gets into a snowplow is:

\be \Dmsnpl = \Ncc \frac{M_{\rm sw}(\Tfin)-M_{\rm int}(\Tfin)}{\Tff}\, .
\label{eq:dmsnpl} \ee

We recall that a fraction \fbo\ (Equation~\ref{eq:fbo}) of the swept
material (both hot and cooled) and of the restored and evaporated mass
is blown-out to the halo.  Defining $\Dmbo =
\Fbo(\Dmev+\Dmrest+\Dmint+\Dmsnpl)$, the system of equations that
describes the mass flows is:

\be \left\{
\begin{array}{lcl}      
\Dmcold &=& \Dminf+\Dmcool-\Dmsfr-\Dmev\\&&+(1-\Fbo)\Dmsnpl \\
\Dmhot  &=& -\Dmcool -\Dmsnpl -\Dmleak -\Fbo\Dmint \\&&+(1-\Fbo)(\Dmev+\Dmrest)\\
\Dmstar &=& \Dmsfr - \Dmrest \\
\Dmhalo &=& -\Dminf + \Dmleak + \Dmbo \\
\end{array} \right. \label{eq:massfluxes} \ee

\noindent
Mass conservation is guaranteed by the condition
$\Dmhot+\Dmcold+\Dmstar+\Dmhalo=0$.

\subsection{Energy flows}

A similar set of equations can be written for the energy flows.  Here
we concentrate only on the energy \eh\ of the hot component, that
determines \th.  The total energy released by SNe is:

\be \Desn = \Ncc \frac{E_{51}\Nsn}{\Tff} \, . \label{eq:desn} \ee

\noindent
The rates of energy loss by cooling, snowplow, blow-out and leak-out
are respectively:

\be \Decool = \frac{\Eh}{\Tcool}
\label{eq:decool} \ee

\be \Desnpl = \Dmsnpl \Th\ \frac{3}{2}\frac{k}{\Muh m_{\rm p}}\, ,
\label{eq:desnpl} \ee

\be \Debo   = \Fbo \Dmint \Th\ \frac{3}{2}\frac{k}{\Muh m_{\rm p}}\, ,
\label{eq:debo} \ee

\be \Deleak = \frac{\Eh}{\Tleak}
\label{eq:deleak} \ee

Regarding the energy budget of the SB (Equation~\ref{eq:eth}), while
thermal energy is obviously given to the ISM kinetic energy is
transformed into turbulence and then partially thermalized.  We have
verified that including or excluding kinetic energy from the energy
budget does not change appreciably the dynamics of the system.  At
present we decide to give it to the ISM, more refined modeling of the
decay of turbulence will be presented elsewhere.

In the blow-out regime the ISM receives a fraction $(1-\Fbo)$ of the
energy of the SB (thermal, Equation~\ref{eq:eth} evaluated at
\tfin\footnote{When \tfin\ is short we force \rsn\tfin\ to be at least
unity.}, plus kinetic) and of the energy connected to the evaporated
mass (that comes from the first SNe exploding in the cloud):

\be \Desb = (1-\Fbo)\left(\Ncc\frac{\Eth+\Ekin}{\Tff}
+\Dmev\Tevap\frac{3}{2}\frac{k}{\Muh m_{\rm p}}\right)
\, . \label{eq:defbbo} \ee

\noindent
In the adiabatic confinement case all energy from SNe is given to the
ISM:

\be \Desb = \Desn\, . \label{eq:defbad} \ee

\noindent
In this case the energy of the evaporated cloud is already included in
the total SN budget.  Finally, in case of PDS confinement the ISM
receives the thermal and kinetic energy of the SB.  As mentioned in
Section 2.7, the energy $E_{\rm rest}$ associated to SNe exploding
after \tfin\ is given either by a fraction \fpds=$0.737(1
-(\Tfin/\Tpds)^{-3.2}) +0.263$ (the case of fast mixing, where
secondary bubbles share the same fate as the principal one; see
Section 2.6) or entirely, \fpds=1 (the case of slow mixing,
where the remaining energy is pumped into the rarefied interior of the
bubble):

\be \Desb = \Ncc\frac{1}{\Tff}(\Eth+\Ekin + \Fpds E_{\rm rest})
\label{eq:defbpds} \ee

\noindent
Given the uncertainty connected to the modeling of $E_{\rm rest}$, we
consider it worthless to include a detailed treatment of the energy
from the evaporated cloud in this case.

The  equation for the evolution of \eh\ is:

\be \Dehot = -\Decool -\Deleak +\Desb -\Debo -\Desnpl
\label{eq:nrgflux} \ee

The efficiency of feedback \fe\ is defined as the thermal energy
gained (or lost) by the ISM (by the hot phase) at the end of the
feedback process, divided by the energy injected by SNe.  Cooling and
leak-out are not directly associated with the action of SBs, while
energy losses by blow-out and snowplow act in decreasing the thermal
energy of the ISM by depleting the hot phase.  We then define \fe\ as:

\be \Fe = \frac{\Desb-\Debo-\Desnpl}{\Desn} \label{eq:fe} \ee

\noindent
It is very important to notice that this quantity is not constrained
to be positive: in particular situations the depleting action of
blow-outs and snowplows can overtake energy injection; in this case
the net effect of SN explosions is a loss of thermal energy more than
a gain.

\subsection{Metal flows}

For each generation of stars a fraction $y$ of the restored mass is
composed by new metals that are continually injected into the ISM.  We
call $M^{Z}_{\rm i}$ (where i=hot, cold, $\star$ or halo) the mass of
metals in the various components, and $Z_{\rm i}=M^{Z}_{\rm i}/M_{\rm
i}$ their metallicities.  In the instantaneous recycling approximation
the system of equations for the metals is:

\be \left\{
\begin{array}{l}
\dot{M}^{Z}_{\rm hot}   = -\Zhot(\Dmcool+\Dmsnpl+\Dmleak+\Fbo\Dmint) \\
\ \ \ \ \ \ \ \ \ \ 
+(1-\Fbo)[\Zcold\Dmev + (\Zcold+y)\Dmrest] \\
\dot{M}^{Z}_{\rm cold}  = \Zhot(\Dmcool+(1-\Fbo)\Dmsnpl) +\Zhalo\Dminf\\
\ \ \ \ \ \ \ \ \ \
-\Zcold(\Dmev+\Dmsfr)\\
\dot{M}^{Z}_\star = \Zcold(\Dmsfr - \Dmrest) \\
\dot{M}^{Z}_{\rm halo}  = -\Zhalo\Dminf+\Zhot[\Dmleak+\Fbo(\Dmint+\Dmsnpl)]\\
\ \ \ \ \ \ \ \ \ \
+\Fbo[\Zcold\Dmev+(\Zcold+y)\Dmrest]\\
\end{array} \right. \label{eq:metalfluxes} \ee

In this case mass is not conserved, the source term being $y\Dmrest$.
In these equations it is implicitly assumed that metals are
efficiently mixed within each component.  This assumption is
reasonable for the hot and cold phases, but is clearly wrong for
stars.  In other words, \zstar\ is the average metallicity of stars
but not that of the last generation, which contributes to enrichment.
As a consequence, we use \zcold\ for the metallicity of the newborn
stars, while the actual distribution of metallicity of stars can
easily be obtained by computing the evolution $\Zcold\Dmstar$ over
time.  Mixing of metals within the halo is another delicate
assumption; if gas is blown out in form of cold clouds then mixing may
be inefficient. A more refined modeling will be required in realistic
cases.

\subsection{Parameters}

It is useful at this point to sum up the parameters introduced in the
model.  Some of them are connected to the theory of stellar evolution
or with the choice of the IMF; we do not regard them as free
parameters of the model.  In this paper we fix their values as
follows: \mstsn=120 \msun, \frest=0.2, \tlife=$2.7\times10^7$ \yr,
$y$=0.04.

The following quantities have been introduced in the various steps of
the feedback model, and should be regarded as free parameters: \alfa,
\shape, \esn, \fstar, \fevap, \tevap, \fcool.  It is worth recalling
that \fevap\ and \tevap\ will be determined in paper II, and that
\tevap\ plays no role if SBs are pressure-confined.  Other parameters
are \fbom, if blow-out is required to be more efficient than the
simple geometrical model of Equation~\ref{eq:fbo}, and \fpds, that
regulates the injection of energy after PDS and is relevant only in
case of PDS confinement.  The parameters \tc\ and $m_{\rm l}$ are
presently kept constant (we have verified that the solutions do not
change much for reasonable changes in these parameters).  This
parameter space is to be considered as minimal: all the computations
presented above are only useful order-of-magnitude estimates, so many
quantities could in principle be fine-tuned (with the aid of new
parameters) to reproduce, for instance, the results of detailed
simulations.

Finally, the total mass \mtot, the volume $V$ (or equivalently the
total density \ro=\mtot/$V$ or surface density \stot=2\heff\ro), the
infall time \tinf, the vertical scale-height \heff, the velocity
dispersion of clouds \sigv\ and the geometry of leak-out $d$ (fixed to
1 in the following) are the parameters connected to the system in
which feedback acts.  These parameters are obviously determined by the
characteristics of the galaxy; in particular, the galaxy-halo system
will never be a ``closed box'' as naively assumed here, so \mtot\ will
not be a constant; \ro\ will be fixed by the gravitational well of the
dark-matter halo and the amount of angular momentum retained by the
gas; \sigv\ will be determined by gravitational infall, dissipation
and re-acceleration by blasts; \heff\ will be determined by \sigv\ and
the surface density of the galaxy; \tinf\ will be determined by the
cooling and infall times of the dark-matter halo.  To avoid the
modeling of the galaxy at this stage, we consider for simplicity these
parameters as independent.  The regions of these parameter space
relevant to galaxies will be determined once their large-scale
structure is fixed.

\section{Results}

The system of Equations \ref{eq:massfluxes}, \ref{eq:nrgflux} and
\ref{eq:metalfluxes} has been integrated with a standard Runge-Kutta
algorithm with adaptive stepsize (Press et al. 1992).  The adaptive
stepsize is computed considering only the equations for the mass and
energy flows.  As mentioned above, the numerical integration is often
delicate, especially when the system switches from one regime to
another.  For this reason it is important to interpolate smoothly
between different regimes.

In the following we fix the total mass of the system to $10^{11}\ {\rm
M}_\odot$.  All the results can be simply rescaled to any total mass,
as long as the mass allows for the presence of at least one collapsing
cloud in the volume.  The initial conditions are set by putting most
gas into the halo, which is both physically reasonable and
computationally convenient.  The volume $V$ is anyway filled with a
small amount of mass in both cold and hot gas and a tiny amount of
stars.  This sets the system into a physically acceptable transient
regime, allowing a smooth integration.  We have verified that in
general the precise choice of these initial conditions does not
influence the result as long as the system starts in a way which is
not pathological.  Moreover, we set all primordial metallicities to
$10^{-4}$.  We specify the density of the system through the quantity
\ro=\mtot/$V$, i.e. the density that the system would have if all the
mass were in the volume $V$ and not in the halo.  Clearly, the actual
density of the ISM will always be smaller than \ro.  For an easier
comparison to astrophysical data, we show results in terms of the
surface density \stot=2\ro\heff.

\subsection{Feedback regimes for a reference choice of parameters}

We choose a reference set of parameters by fixing them to the typical
(or tentative) values quoted above in the text: \alfa=2, \esn=1,
\fstar=0.1, \fcool=0.1, \shape=0.2, \fevap=0.1, \tevap=$10^6\
K$.  For this set of parameters we run the system of equations for a
grid of values in the \heff--\ro\ (or equivalently \heff--\stot)
plane, with a time scale of infall \tinf=$10^9$ yr, which is
suggested for the Milky Way by galaxy evolution models (see, e.g.,
Chiappini, Matteucci \& Romano 1997), and a velocity dispersion of
clouds \sigv=10 \kms, typical of spiral discs.  We stop the
integration at three infall times, and check the actual regime of
feedback.  We have verified that the feedback regime may change with
time, but in most cases it does not change from 1 to 3\tinf.

\begin{figure*}
\centerline{
\includegraphics[width=8.5cm]{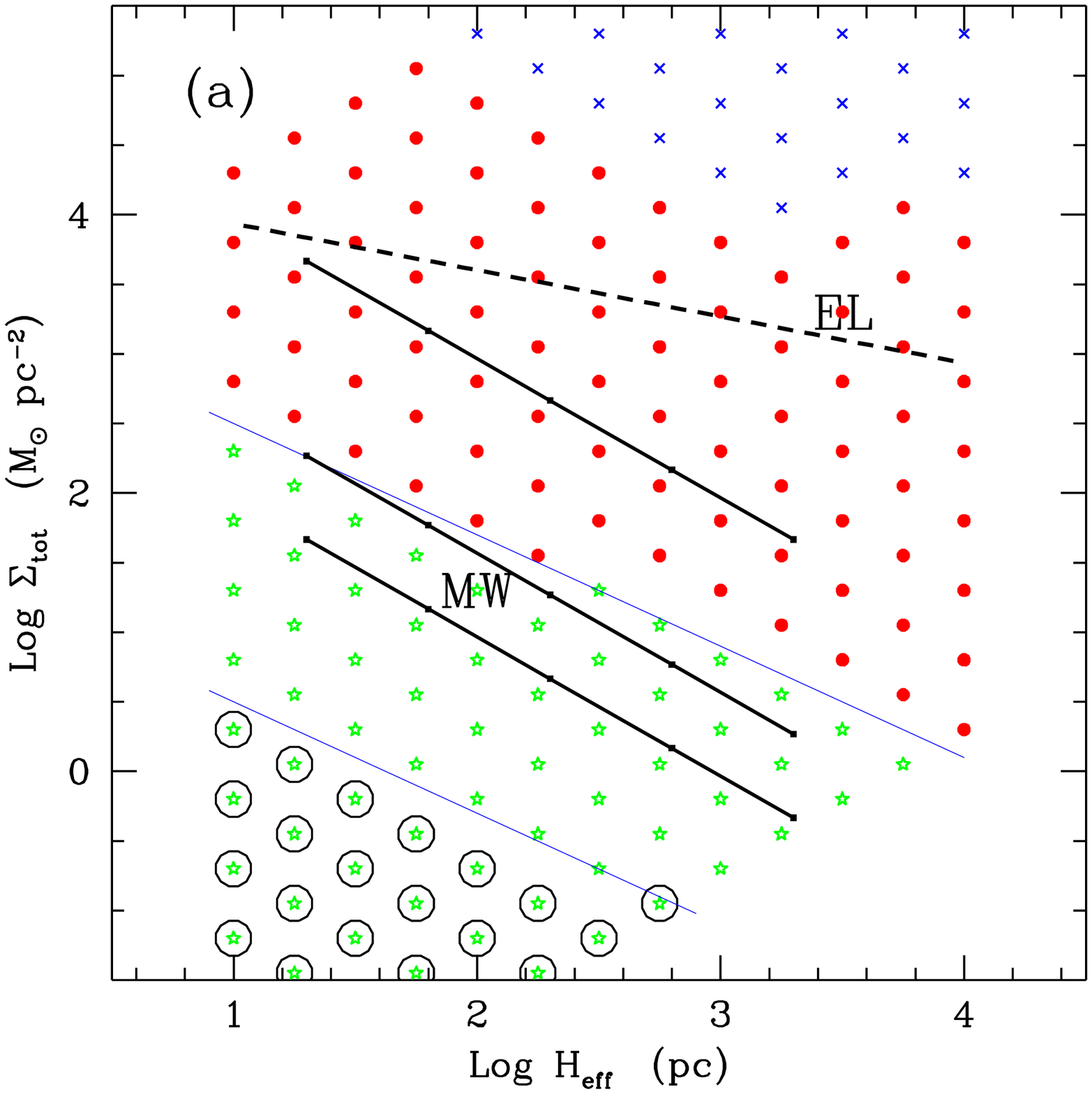}
\includegraphics[width=8.5cm]{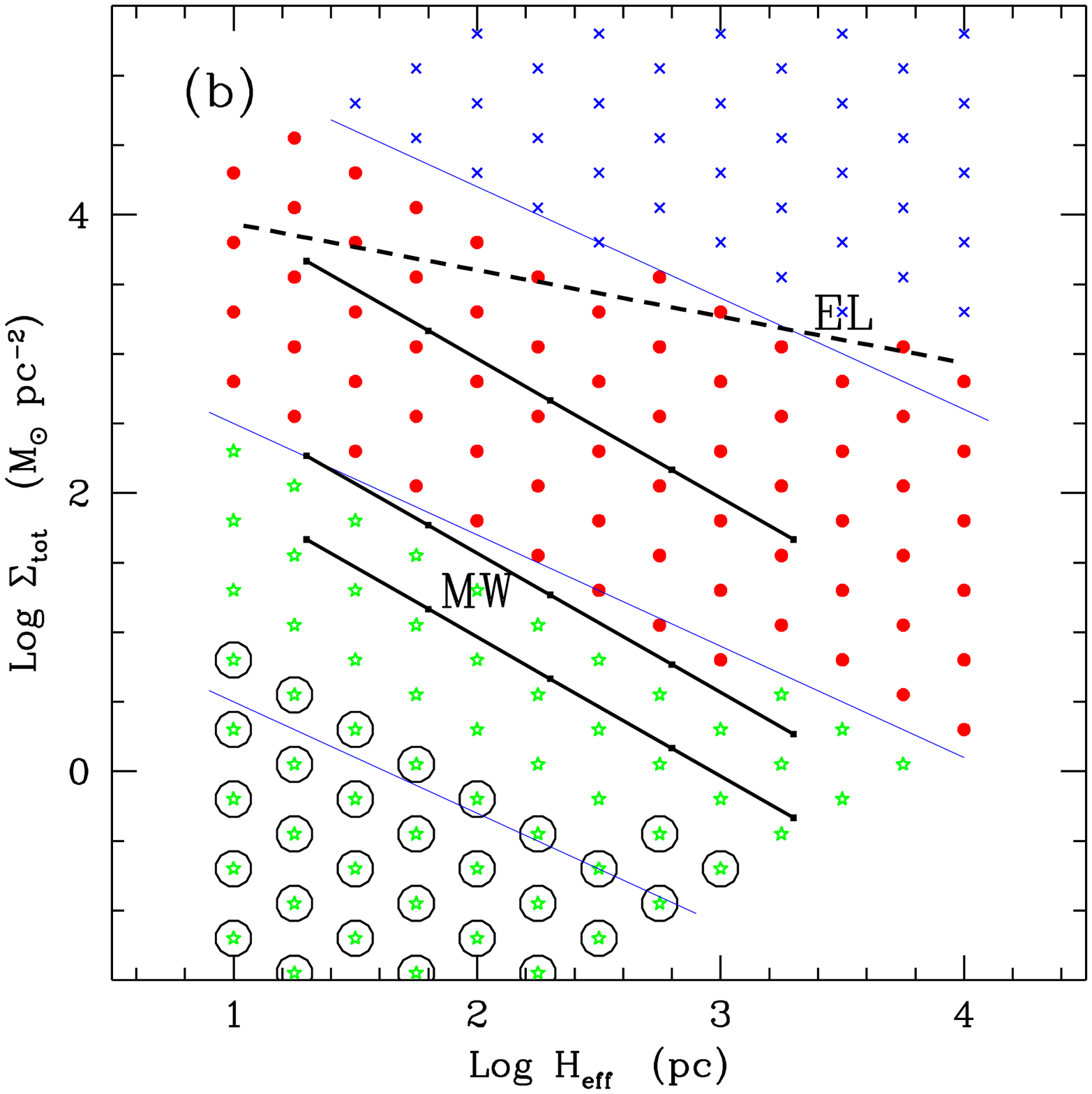}}
\caption{Feedback regimes in the \heff--\stot\ plane after three
infall times.  Filled circles denote adiabatic confinement, stars
adiabatic blow-out, crosses PDS confinement.  Critical cases are
marked by circles.  Thin lines separate regions with different regimes
(see Equation~\ref{eq:limit_bo}), thick continuous lines show the position
of spiral discs with (starting from the lowest) \sigv=5, 10 or 50
\kms, while the thick dashed line shows the average position of
bulges.  The labels MW and EL are relative to the examples shown
below.  Panel (a): reference choice of parameters.  Panel (b):
\esn=0.3 and \fevap=0.05.}
\label{fig:ciclo}
\end{figure*}

Fig.~\ref{fig:ciclo}a reports the regions in the \heff--\stot\
plane in which various feedback regimes are found.  At low \stot\ and
\heff\ values the SBs are able to blow out, and when they do they are
always in the adiabatic stage.  For increasing \heff\ it is more and
more difficult to blow out, so that SBs are kept confined by the
external pressure in the adiabatic stage.  The two regimes are roughly
separated by the relation shown in the figure:

\be \Stot = 8 \left(\frac{\Heff}{1000\ {\rm pc}}\right)^{-0.8}
\ \ {\rm M}_\odot\ {\rm pc}^{-2} \label{eq:limit_bo} \ee

At densities lower than this limit by roughly two orders of magnitude
the system gets into a critical behaviour, where the hot phase is
strongly depleted and the filling factor of the cold phase becomes
high. This regime will be described in Section 4.3.  At very high
surface densities PDS confinement is met; this will be discussed
later.

If the vertical scale-length of a disc is set by dynamical equilibrium
then $\Heff=\sigma_{\rm v}^2/\pi G\Stot$.  Galaxy discs will then lie
on the continuous lines shown in Fig. \ref{fig:ciclo} for \sigv=5, 10
and 50 \kms\footnote{Here we have assumed that the vertical
scale-length of the molecular clouds is \heff/2, which amounts to
halving \heff\ to take into account the easier blow-out of SBs that
are off the mid-plane; see the discussion in Section 2.7}.  Discs with
the canonical of \sigv=7 \kms\ will be in the adiabatic blow-out
regime, except possibly in the inner parts (where they blend with
bulges), while discs with \sigv$>$10 \kms\ will be in the adiabatic
confinement regime.  Bright spheroidal galaxies roughly follow a
relation of the kind $R_e=22 (M_{\rm tot}/10^{12}\ {\rm
M}_\odot)^{0.6}$ \kpc\ (proposed by Chiosi \& Carraro 2002), that can
be extrapolated to meet the locus of globular clusters.  Identifying
the effective radius $R_e$ with \heff, this relation is shown in the
\heff--\stot\ plane as a dashed line.  It is clear that feedback in a
spheroid will typically be in the adiabatic confinement regime.

\begin{figure*}
\centerline{
\includegraphics[width=18cm]{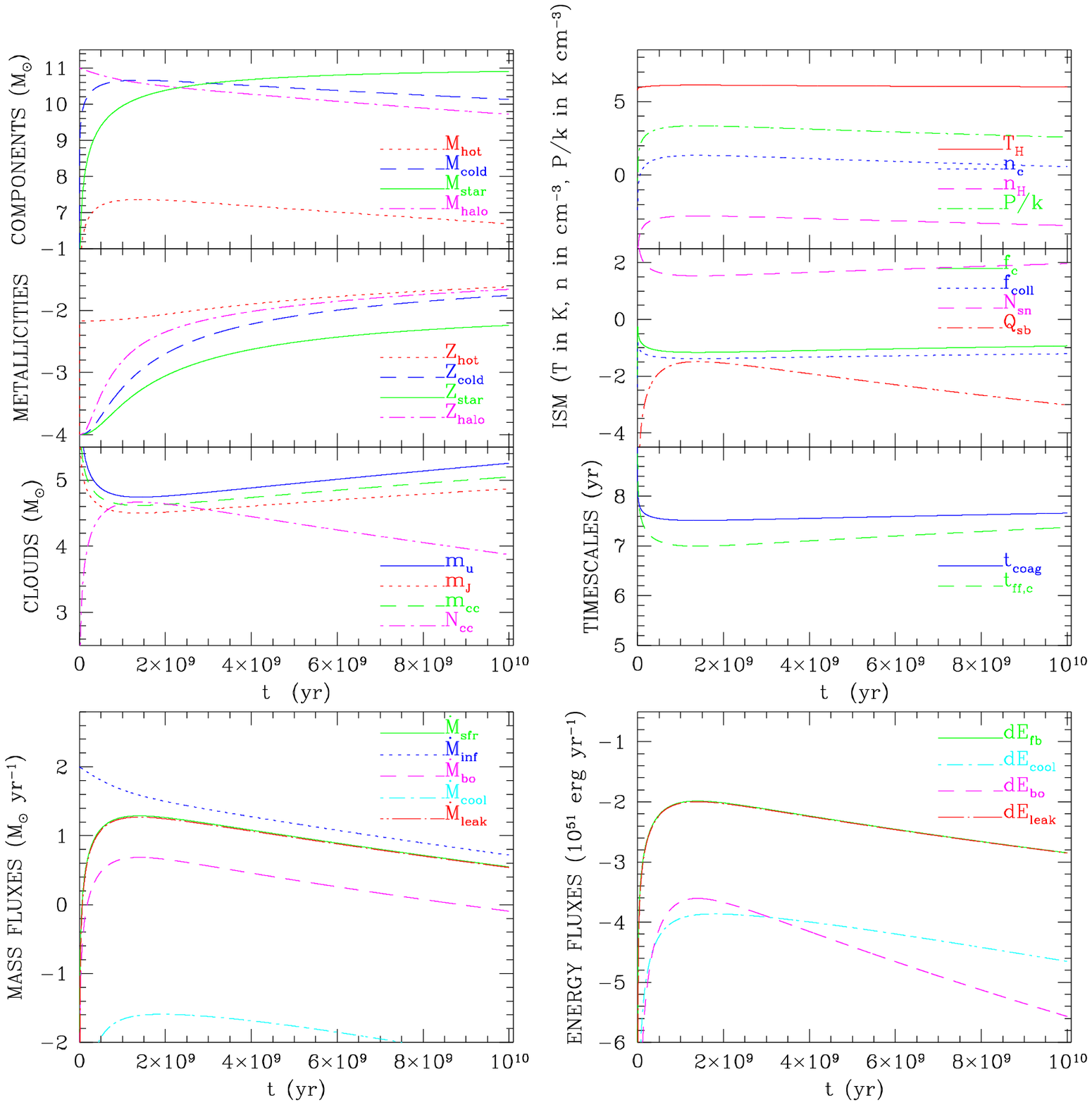}
}
\caption{Evolution of the system for the MW adiabatic blow-out example
with (\heff, \ro)=(100 \pc, 0.1 \dens), \stot=20\surf.  The panels
show the main properties of the systems, the quantities and their
units are given in the labels.  Time is linear, all the quantities
given in ordinate are logarithmic.  In the lower panels of mass and
energy flows, the \dmsfr\ and \dmleak\ curves and the \desb\ and
\deleak\ curves are very similar and hardly distinguishable.}
\label{fig:stadbo}
\end{figure*}

\begin{figure*}
\centerline{
\includegraphics[width=18cm]{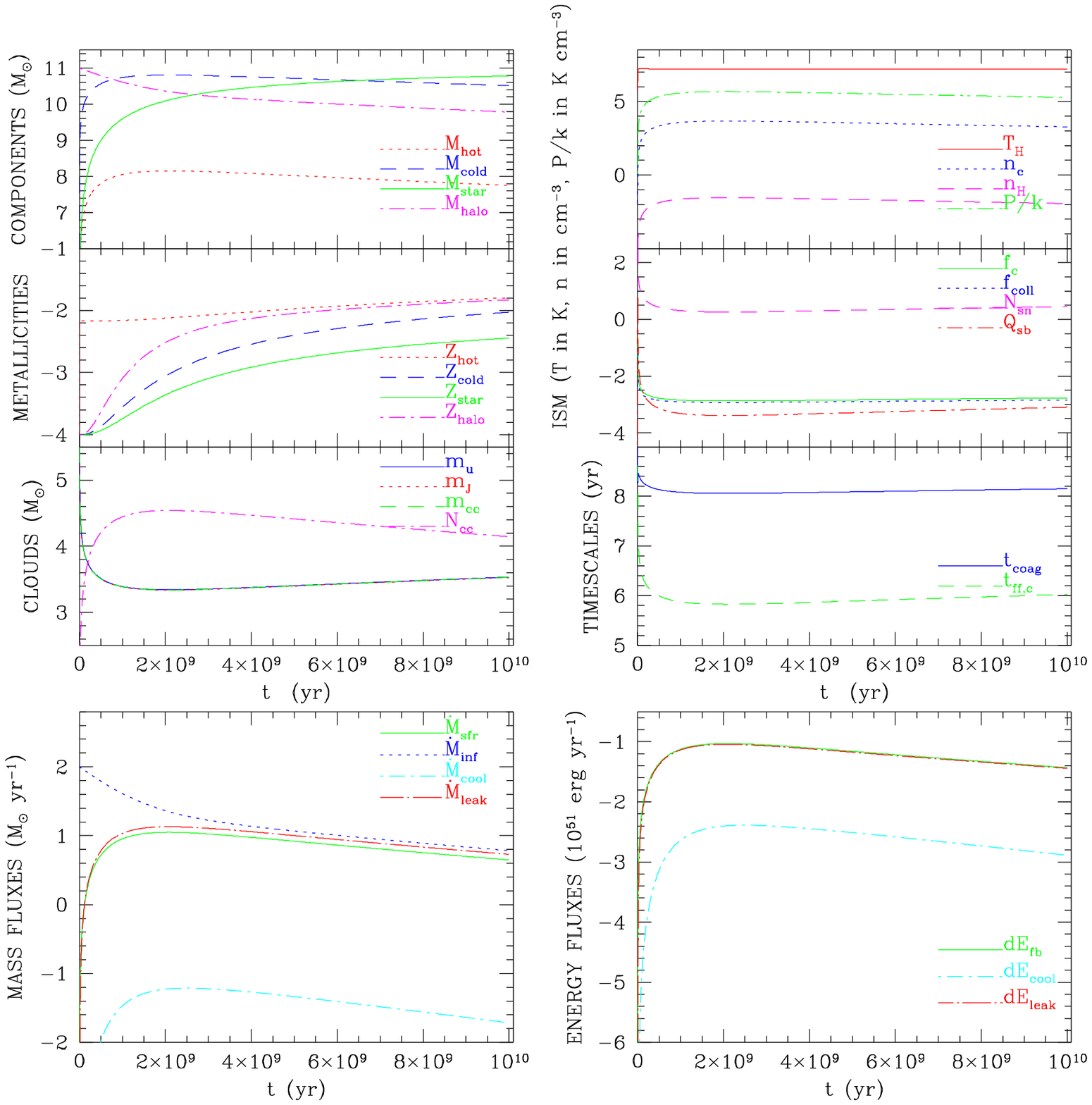}
}
\caption{As in Fig.~\ref{fig:stadbo}, for the EL adiabatic confinement
example with (\heff, \ro) = (3 \kpc, 0.3\dens) or \stot=1800 \surf.}
\label{fig:stadcn}
\end{figure*}

\begin{figure*}
\centerline{
\includegraphics[width=18cm]{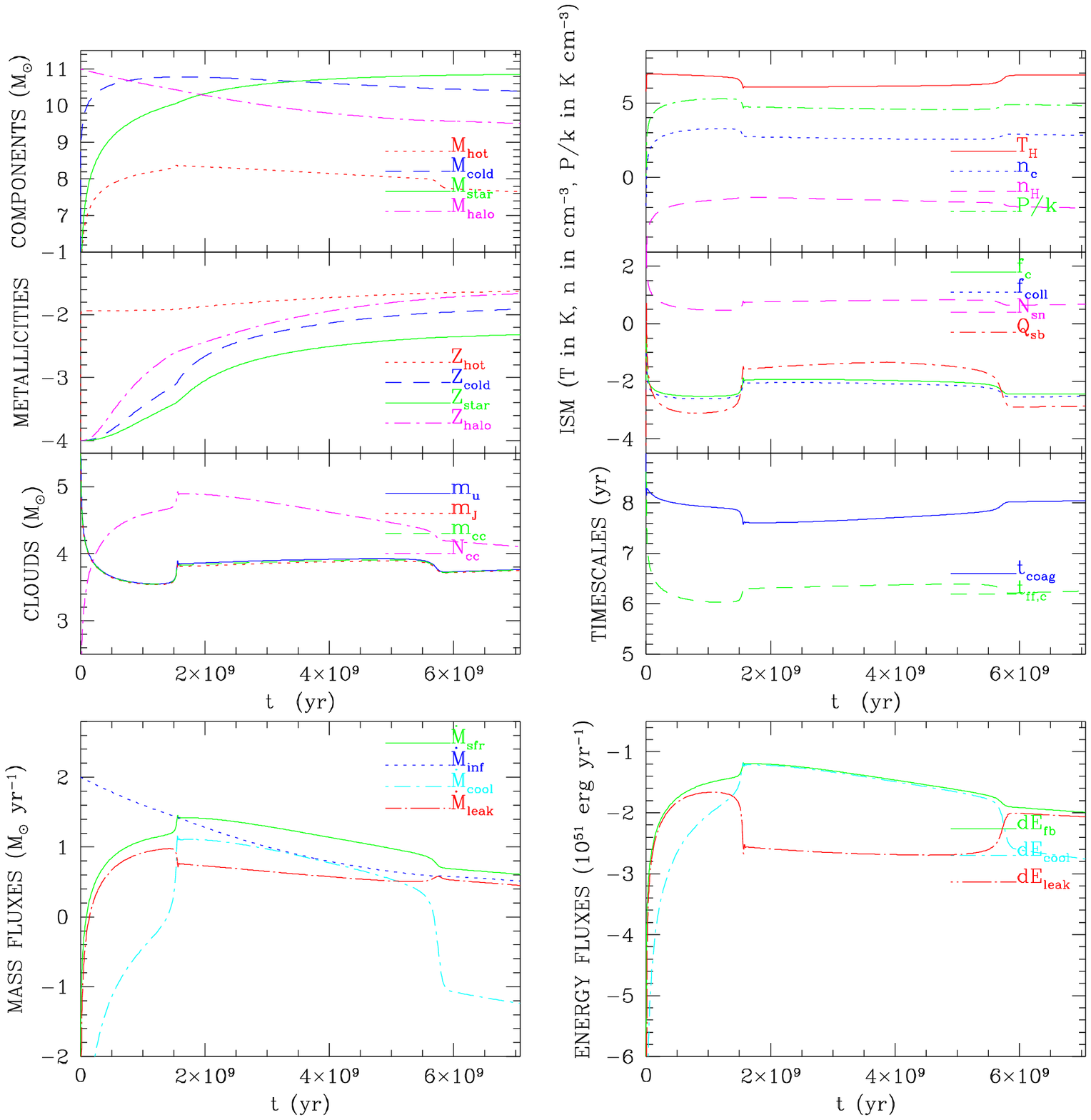}
}
\caption{As in Fig.~\ref{fig:stadbo}, for the EL PDS confinement
example (\esn=0.3, \fevap=0.05, \fpds=1).}
\label{fig:pdscon}
\end{figure*}

\subsection{Some examples}

We show here examples of the evolution of the system in various
regimes.  In particular we show the evolution, up to 10 \Gyr, of
masses and metals of the four components, mass and energy flows, ISM
and cloud properties.  The evaporation and restoration rates, \dmev\
and \dmrest, are not shown in the figures as they are simply
proportional to (and smaller than) the star-formation rate \dmsfr.

Fig.~\ref{fig:stadbo} shows a Milky Way-like system in the adiabatic
blow-out regime, with (\heff, \ro)=(100 \pc, 0.1 \dens) or \stot=20
\surf\ (it is denoted as MW in Fig.~\ref{fig:ciclo}).  Gas cools but
does not transform promptly into stars; in fact, it is continually
recycled into collapsing clouds.  The final \mcold/\mstar\ ratio is
\circa0.1.  The hot phase regulates to a fraction $F_{\rm
h}$\circa$2\times 10^{-4}$.  The halo gas is continually recycled by
infall, blow-out and leak-out.  Regarding metals, the hot phase is
promptly enriched to nearly solar values, followed by the halo gas,
which receives the blown-out and leaked-out metals; the final
metallicity of the cold gas is solar, those of hot and halo components
are 60 per cent higher.  The star formation rate \dmsfr\ after a rise
of \circa1 \Gyr\ decreases exponentially from \circa20 to \circa3
\msunyr\ in \circa9 \Gyr; the average value of the star-formation rate
is \circa10 \msunyr.  The infall rate \dminf\ after \circa2 \Gyr\
regulates to slightly higher values, while the leak-out rate \dmleak\
is very similar to the star-formation rate.  Notably, leak-out
dominates over blow-out, and the cooling term \dmcool\ is very low.
The energy equation is characterized by the near equality of the two
main flows, \desb\ and \deleak; this reflects in the very stable value
of \th\circa $10^6$ \K.  The ISM is self-regulated and weakly varying
over many infall times; its properties are \nh\circa $10^{-3}$ \cmt,
\nc\circa 10 \cmt, $P/k$\circa $10^3$ \K\ \cmt, and change by a factor
\circa5 from 1 to 10 \Gyr.  The porosity \poro\ is low, indicating
that active bubbles (i.e. expanding blasts) do not dominate the
volume.  However, if we assume that SBs are recognizable for a time
\tlife\ (see Section 2.7), the porosity of ``observed'' SBs results as
high as \circa1 at a few \Gyr\ and \circa0.1 at the end of the
integration; this implies an apparently bubble-dominated ISM.  The
filling factor \fc\ of the cold phase is \circa0.1, while the fraction
\fcoll\ of cold gas in collapsing clouds is slightly lower.  The
population of collapsing clouds is also rather stable, with masses in
a range of roughly a factor of two around \mcc\circa$10^5$ \msun.  The
coagulation time \tcg\ is higher than the dynamical time \tff by a
factor of a few, and is \circa$5\times10^7$ yr, not very different
from the dynamical time of the Milky Way disc.  Finally, both the
number of clouds and the number of SNe per cloud are high enough to
justify the assumptions of the model.

Fig.~\ref{fig:stadcn} shows an example of adiabatic confinement (EL)
that lies near the elliptical line (Fig.~\ref{fig:ciclo}), with
(\heff, \ro) = (3 \kpc, 0.3\dens) or \stot=1800 \surf. In this case,
despite of the higher density, gas is consumed more slowly than the
previous case, and the final fraction of \mcold/\mstar\ is still
\circa0.5.  The fraction of hot to cold gas is $F_{\rm
h}$\circa$10^{-3}$.  The pattern of chemical enrichment is similar to
the previous cases, although metallicities are lower at the final time
(due to the lower amount of gas consumed).  Mass flows peak to
slightly lower values, and decrease less steeply at later times.
Again, star formation regulates nearly to the infall rate after a
few infall times, and leak-out is only slightly lower.  Blow-out is
obviously absent and cooling is again negligible.  As before, feedback
and leak-out energy flows nearly compensate each other.  The ISM is
characterized by high pressure ($P/k$\circa $10^5$ \press), density of
both phases (\nc\circa$10^3$ \cmt, \nh\circa$10^{-2}$ \cmt),
temperature of the hot phase (\th\circa$1.5\times10^7$) and a
correspondingly lower filling factor of the cold phase (\circa$2\times
10^{-3}$).  Collapsing clouds are very small (\mcc\circa$3\times
10^3$\msun), and this is due to the very high density of the cold
phase with the consequent low Jeans mass.  Besides, the range of
collapsing masses is tiny due to the low dynamical time and the
consequent inefficient coagulation.  This reflects into a low fraction
of cold mass in collapsing clouds (\fcoll\circa$10^{-3}$) and a low
porosity of active SBs.

This example is useful to understand the change in the behaviour of
the system from the adiabatic blow-out to the adiabatic confinement
regimes, but clearly a spheroid forms on shorter time-scales than 1
Gyr, and this leads (within the same physical time) to a consistently
more rapid star formation, higher star-formation rates, higher
enrichment, higher pressure and densities of ISM, slightly smaller
collapsing clouds; however, the temperature of the hot phase and the
filling factor and porosity of the cold phase are rather insensitive
to the infall time.  A similar trend is observed when the density is
increased at fixed \heff.

The very low number of SNe per cloud in the adiabatic confinement case
highlights a limit of applicability of the model in this case.
However, it is easy to check that in the adiabatic confinement regime
all terms in the system of Equations (\ref{eq:massfluxes},
\ref{eq:nrgflux} and \ref{eq:metalfluxes}) are independent of the
actual size of SBs.  The thermal energy of the first blast will be
radiated away before the SNR manages to destroy the star-forming
cloud, so a lower effective value of \esn\ will be reasonably used.
The high value of the density of the cold phase highlights another
problem.  The reason why cold gas is not promptly consumed into stars
is that it waits to be included in collapsing clouds.  But for such
high \nc\ values the assumption that gravitational collapse is
required to trigger the formation of H$_2$ is probably wrong, and star
formation is likely to be spread throughout the cold phase.  This can
be reproduced simply by forcing \fcoll\ to be unity; in this case the
evolution of the system becomes trivial, the main mass flows (\dmsfr,
\dminf, \dmleak\ and \dmev) become all proportional to each other and
decay exponentially over one infall time.

As shown in Appendix C (Fig.~\ref{fig:plane2}), to reach the PDS stage
at \th=$10^6\ K$ it is necessary to have rather high densities \nh\
and relatively high mechanical luminosities \lmech; the constraint
tightens considerably at higher temperature.  At high densities the
Jeans mass is rather low, then \lmech\ values are much smaller than
unity, so that SBs are mostly kept confined in the adiabatic stage.
Higher mechanical luminosities could be achieved by increasing \esn,
but this implies also a higher \th.  As shown in Table C1, the ratio
between \tpds\ and \tconf\ is proportional to
$L_{38}^{-5/22}\Th^{5/4}$; as a consequence, the advantage in
decreasing \lmech\ is over-compensated by the increase in \th; as a
consequence, PDS confinement is more easily achieved by lowering \esn.
We recall that for small clouds an effective lower \esn\ value is
reasonably obtained because the thermal energy of the first blast is
lost before the blast gets out of the collapsing cloud; moreover, for
such dense clouds a lower value of \fevap\ is likely (paper II).
Fig.~\ref{fig:ciclo}b shows the regimes in the \heff--\stot\ plane for
\esn=0.3 and \fevap=0.05; while the limit between adiabatic blow-out
and confinement (fig.~\ref{fig:ciclo}a) is unchanged (and critical
cases are found at densities higher by a factor 3), at densities
roughly higher than

\be \Stot = 2500\ (\Heff/1000\ {\rm pc})^{-0.8}\ {\rm M}_\odot\ {\rm
pc}^{-2}  \label{eq:limit_pds} \ee

\noindent
PDS confinement is achieved.  The reason why it takes place
preferentially at high \heff\ values is because leak-out is
inefficient in depleting the hot phase and thus \nh\ is higher.  PDS
confinement solutions are found also in Fig.~\ref{fig:ciclo}a, though
at very high surface densities.

The evolution of the system depends sensitively on how the energy of
SNe exploding after \tfin\ is given to the ISM.  If mixing of hot
phase and hot SB gas is slow, energy is pumped efficiently into the
hot rarefied medium of the stalled bubble; this corresponds to
\fpds=1.  Fig.~\ref{fig:pdscon} shows again the EL example of
fig.~\ref{fig:stadcn} in this case.  PDS confinement starts after a
period of \circa1.5 infall times of adiabatic confinement.  Due to the
sudden lower injection of energy into the ISM, \th\ decreases by
nearly an order of magnitude at the start of PDS confinement.  This
cooling has the effect of increasing \tpds; the system then
self-regulates to a configuration in which SBs stop just after PDS, so
that the shell never acquires much mass.  Pressure and densities are
still high, but the filling factor of the cold phase is as high as
\circa$10^{-2}$.  As a consequence collapsing clouds are bigger, and
\fcoll\ is higher.  At the onset of PDS confinement the star-formation
rate jumps to a value of \circa30 \msunyr, and then decreases slowly.
Due to the lower \th, the fraction of hot gas is slightly high,
$F_{\rm h}$\circa$10^{-3}$.  In this regime cooling is more important
than leak-out in terms of both mass and energy flows; we have
verified that this is always the case for high \heff\ and \stot\
values.  On the other hand, the snowplow flows are small (they are
below the range of the energy flux panel), indicating that the most
relevant effect of PDS is on the structure of the ISM more than on
the mass flows.  Notably, the porosity of the SBs increases by more
than an order of magnitude.  At \circa6 Gyr the ISM amounts roughly to
40 per cent of the total mass, and the solution switches back to
adiabatic confinement.

For smaller infall times or higher densities the PDS regime is
triggered at an earlier time; again we have consistently higher
star-formation rates, metallicities, pressure and densities of the
ISM, while \th, \fcoll\ and \poro\ are hardly affected.

\begin{figure*}
\centerline{
\includegraphics[width=9cm]{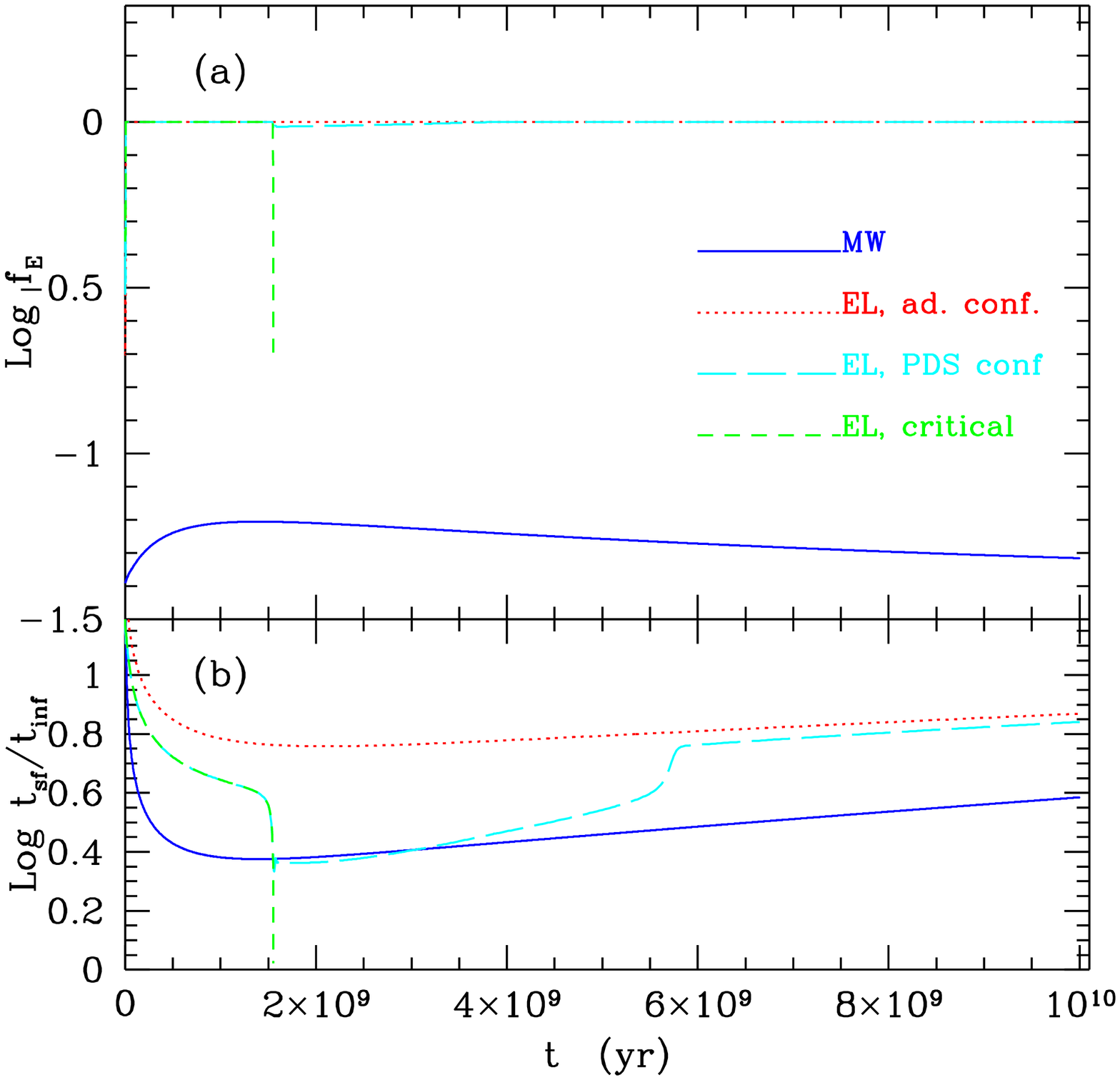}
\includegraphics[width=9cm]{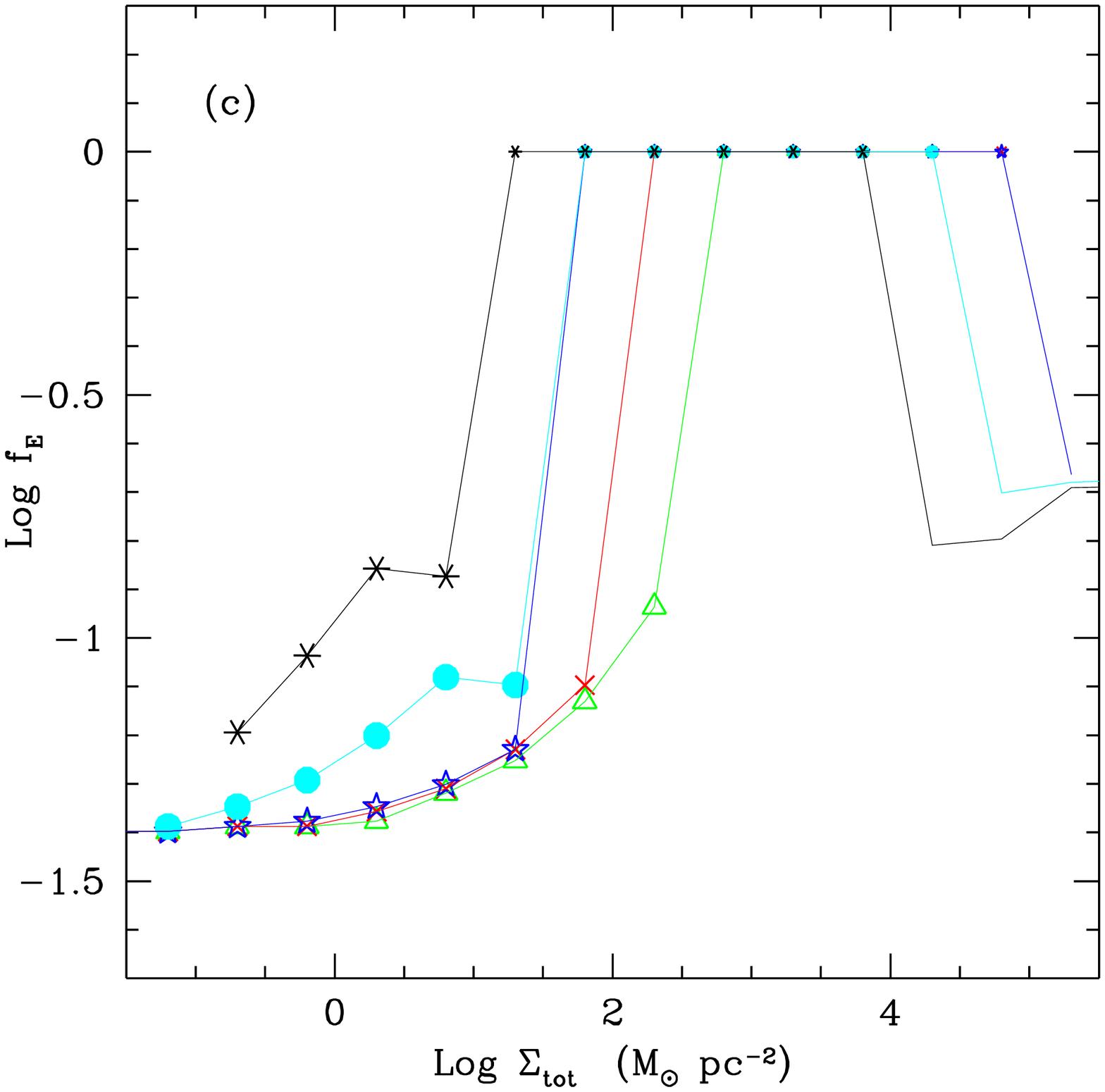}
}
\centerline{
\includegraphics[width=9cm]{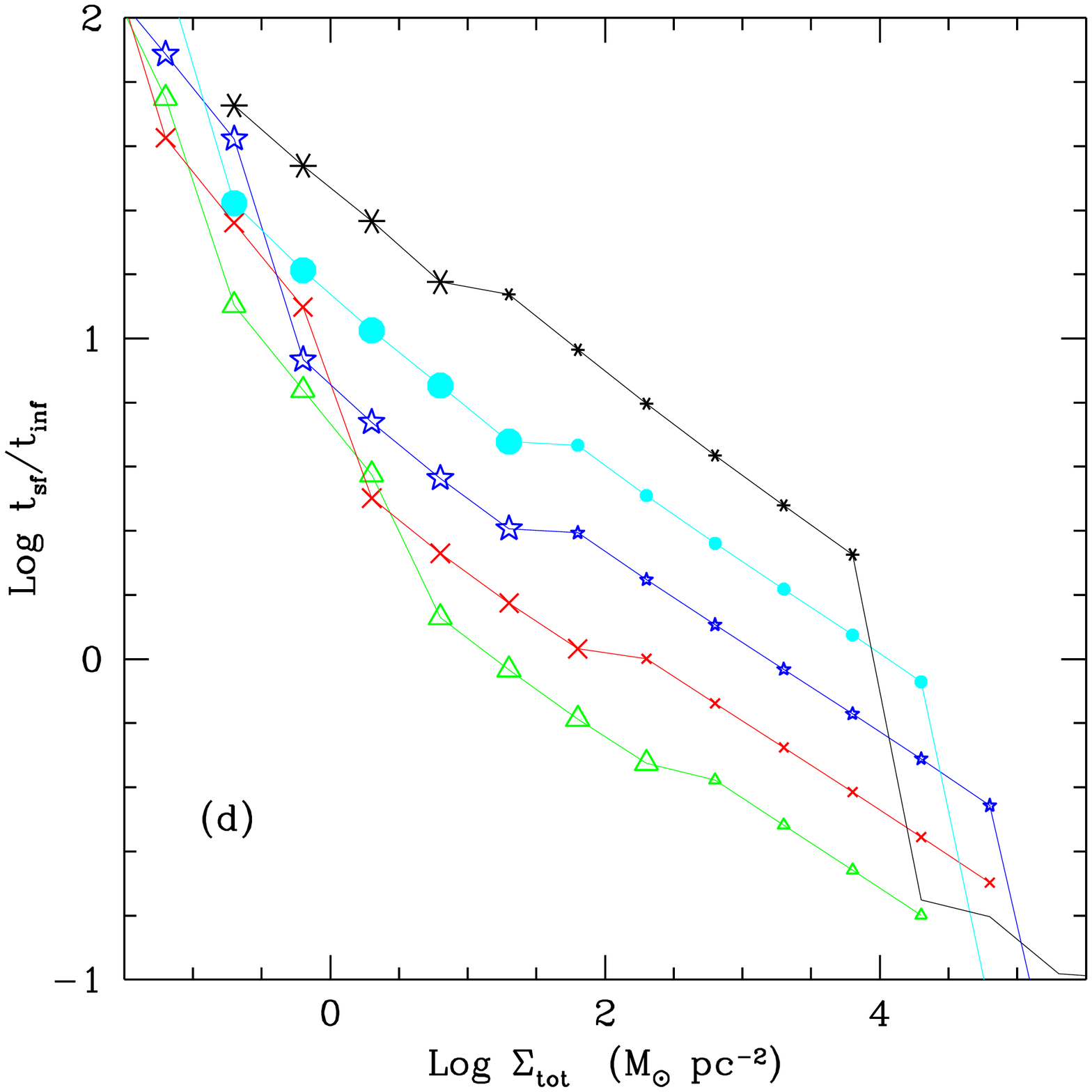}
\includegraphics[width=9cm]{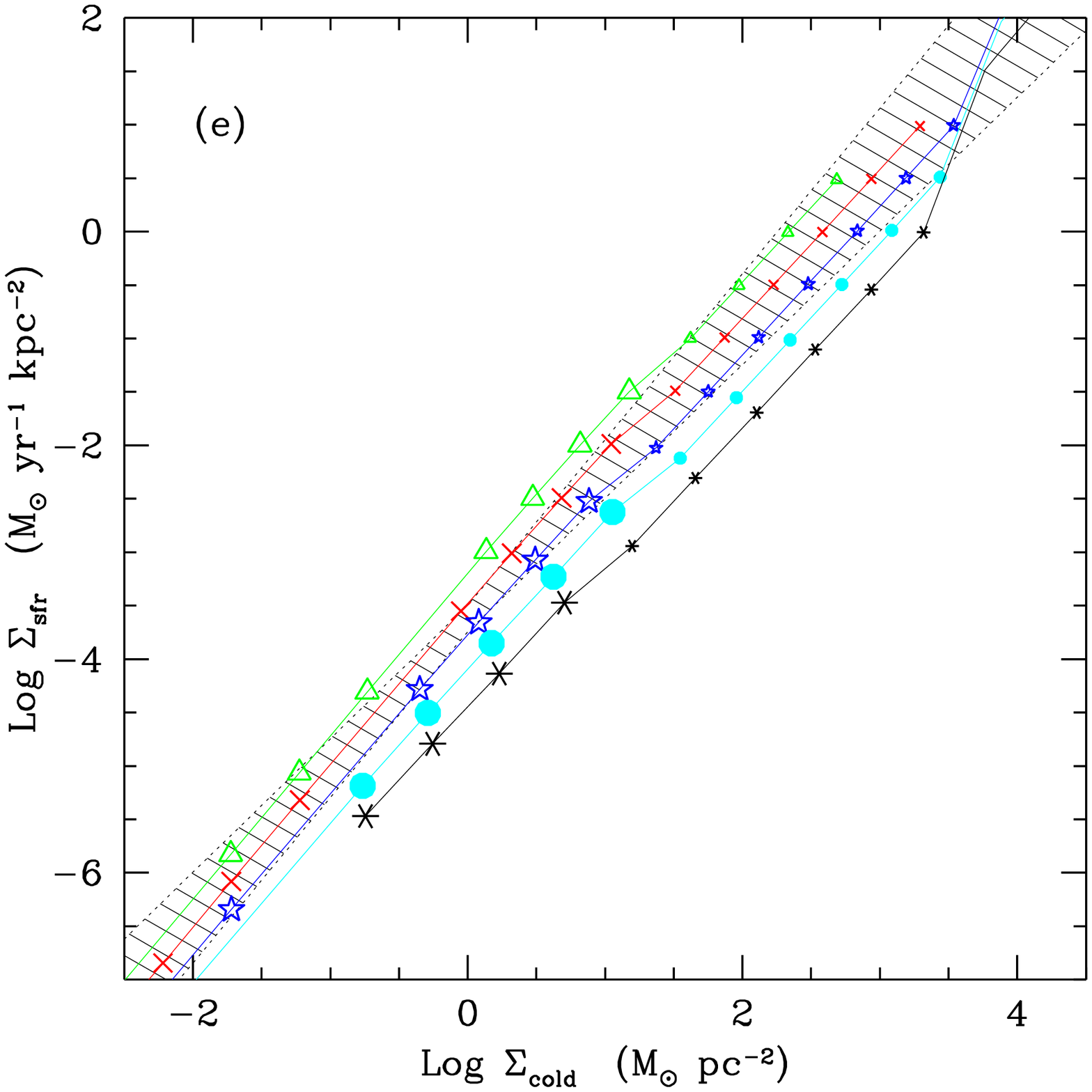}
}
\caption{(a) Efficiency of feedback \fe\ as a function of time for the
examples MW and EL with adiabatic or PDS (self-regulated or critical)
confinement. (b) Star-formation time-scales (in units of the infall
time) for the four cases of panel (a). (c) \fe\ at three infall times
(or at the onset of critical behaviour) for the cases shown in
Fig.~\ref{fig:ciclo}a (with the standard choice of parameters);
triangles, crosses, stars, circles and asterisks are relative to
\heff=10, 30, 100, 300 and 1000 pc; larger points denote the adiabatic
blow-out regime.  (d) Relation between star-formation time-scale and
surface density; symbols as above. (e) Prediction of the Schmidt law
versus the observational relation of Kennicut (1989); symbols as
above.}
\label{fig:eff}
\end{figure*}

Fig.~\ref{fig:eff}a shows the efficiencies of feedback \fe\
(Equation~\ref{eq:fe}) in the examples discussed above.  Unity values are
obtained in the adiabatic confinement regime, while in the PDS
confinement example the efficiency is just a few percent below unity
(as apparent in panel (c) smaller values like 0.5 are obtained at
higher densities or lower infall times).  In the adiabatic blow-out
regime most energy is lost to the halo and the efficiency is slightly
lower than 0.1.  Another way to quantify the efficiency of feedback is
through its effect on star formation.  We quantify it by the ratio
between the star formation time \tsf\ (Equation~\ref{eq:tsf}) and the
infall time \tinf.  As shown in Fig.~\ref{fig:eff}b, a higher
efficiency corresponds to a longer star-formation time scale, but the
correspondence is only qualitative.  For instance, the EL PDS and
adiabatic confinement cases have very similar feedback efficiencies
but, due to the different ISM structure, star-formation time-scales
that differ by more than a factor of two.

It must be kept in mind that \fe\ refers to the efficiency with which
the energy of SNe is given to the ISM.  In the case of adiabatic
blow-out, while \circa5-10 of the energy is given to the ISM, \circa5
per cent is lost in the destruction of the star-forming cloud (see
paper II), and a comparable amount is likely lost in the acceleration
of the bubble at blow-out.  The remaining \circa80 per cent of the
budget will be available to heat up the halo gas.

Fig.~\ref{fig:eff}c and d show \fe\ and \tsf/\tinf\ at three infall
times (3 Gyr, or at the final time in critical cases) for the grid of
models shown in Fig.~\ref{fig:ciclo}a.  The efficiency of feedback
\fe\ jumps from a value 0.05--0.1 in the adiabatic blow-out cases to 1
in the adiabatic confinement cases and then down to \circa0.3 in the
PDS confinement cases.  The star-formation time-scale \tsf\ is roughly
fit by a relation:

\be \Tsf=25\, \Tinf \left(\frac{\Stot}{1\ {\rm M}_\odot\ {\rm pc}^{-2}}\right)^{-0.3}
\left(\frac{\Heff}{1\ {\rm kpc}}\right)^{0.5}\, . \ee

\noindent
Critical and PDS confinement cases fall out of this relation.

In spiral galaxies the star formation rate is well correlated with the
amount of cold gas, following the so-called Schmidt (1959) law,
quantified by Kennicut (1989) as $\Sigma_{\rm sf}= (2.5\pm0.7)
\times10^{-4} \Sigma_{\rm cold}^{1.4\pm0.15}\ {\rm M}_\odot\ {\rm
yr}^{-1}$ kpc$^{-2}$ (with $\Sigma_{\rm cold}$ the surface density of
cold gas in \surf).  Fig.~\ref{fig:eff}e shows the predictions of
this relation for the same grid of models at 3 Gyr, compared to the
Kennicut relation.  While the slope is accurately reproduced, the
normalization depends on \heff, and is well reproduced for
\heff\circa50-100 \pc.  Bright spirals are known to have roughly constant
surface densities and velocity dispersion of clouds, so the average
value of \heff\ is also constant and of order of its value in the
solar neighbourhood.  So, the predictions of this model satisfy the
Schmidt-Kennicut law also in its normalization.  We have verified that
this agreement holds in a very broad range of cases and at all times.
This implies that this is a robust prediction of the model, but cannot
be used to fine-tune the parameters.  The Schmidt-Kennicut law is
naturally obtained if star formation depends on the mass of the cold
gas divided by its dynamical time.  In our case
(Equation~\ref{eq:dmsfr}) the relation is not built-in, due to the
presence of the \fcoll\ fraction and to the fact that the dynamical
time is computed on the actual and not average density of the cold
phase.  The Schmidt-like law then follows from the approximate
constancy of \fcoll\ and \fc.

Finally, the Schmidt-Kennicut law is not followed in the external
regions of spiral galaxies, where star formation is quenched.  This is
not predicted by the present model.  However, such star-formation
edges are usually though to be an effect of differential rotation or,
according to Schaye (2004), of photo-heating by the cosmological UV
background.  Both processes are not included here, so this
disagreement is expected and is not considered as a worry.

\subsection{Critical examples}

As long as the system of equations described in Section 3 holds, the
physical system is self-regulated; as we have seen, equilibrium
solutions are found in which the ISM is relatively stable for many
infall times.  However, there are critical cases where the conditions
for the existence of a two-phase medium are violated and the system of
equations does not hold any more.

At densities roughly two orders of magnitude lower than the limit
shown by Equation~\ref{eq:limit_bo} the system gets into a critical regime,
where the hot phase is strongly depleted and the filling factor of the
cold phase \fc\ is larger than that of the hot phase \fh, violating
the assumption of a cold phase fragmented into well-separated clouds.
The cases where this happens are highlighted by a circle in
Fig.~\ref{fig:ciclo}.  The reason for this behaviour is simple: at
such low densities star formation is very weak, while for a thin
structure leak-out is strong, so the hot phase cannot be sustained.
This is not very informative, as such low-density thin systems, if
they exist, would be kept ionized by the cosmological UV background,
so star formation would never start.

\begin{figure*}
\centerline{
\includegraphics[width=18cm]{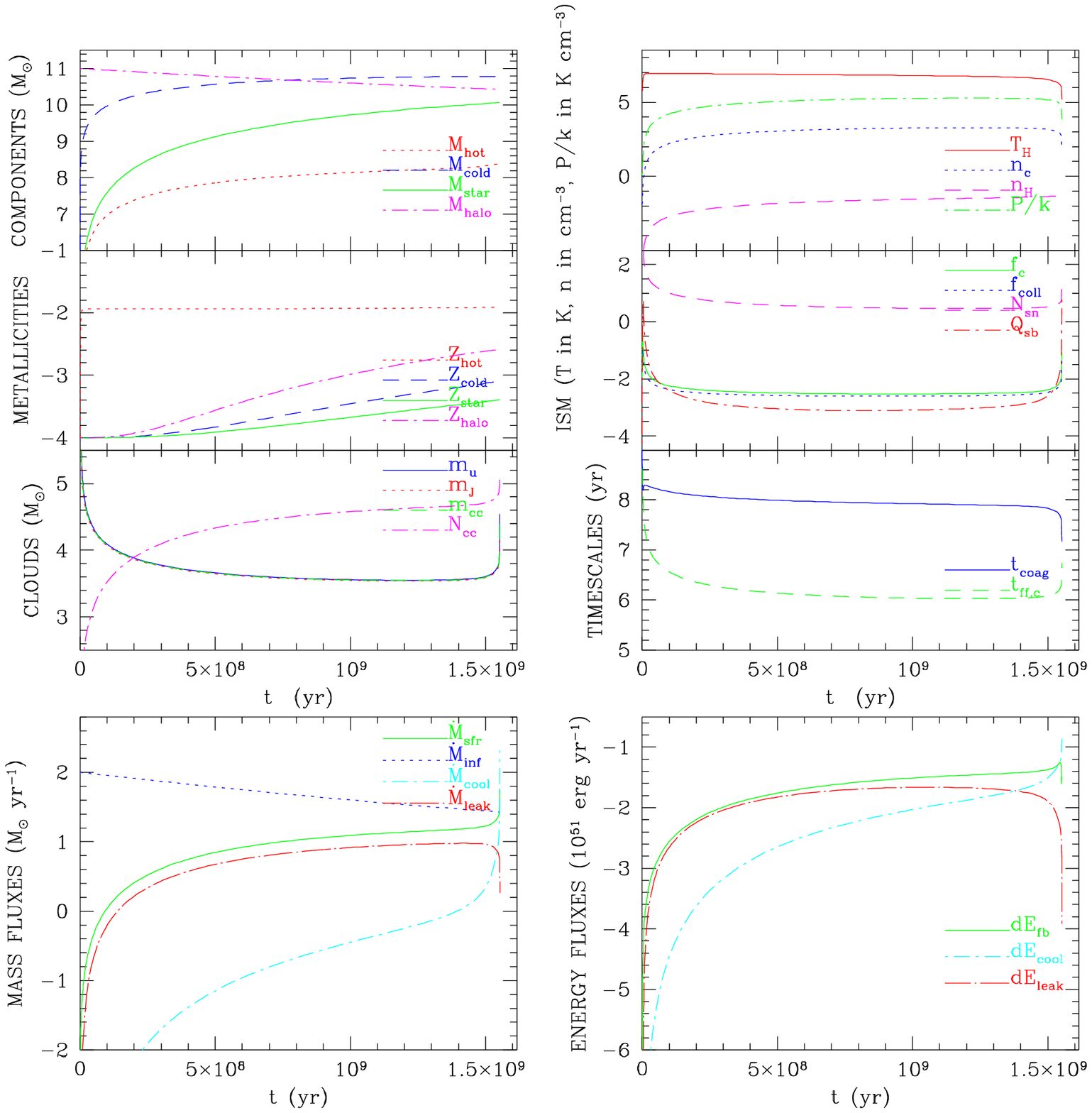}
}
\caption{As in Fig.~\ref{fig:pdscon}, but with \fpds=$0.737(1
-(\Tfin/\Tpds)^{-3.2}) +0.263$.}
\label{fig:pdsgno}
\end{figure*}

A similar phenomenon happens in other cases for different reasonable
choices of the parameters, with the difference that high values of
\fc\ are obtained not from the start but after some time.  If we allow
$f_{\rm bo,max}$ to be as large as 0.8, the blow-out mass flux is much
stronger and dominates over leak-out.  For \stot\circa200--1000 \surf,
when \heff\ increases above \circa1 \kpc\ the system does not go into
the adiabatic confinement regime; SBs get bigger and blow-out gets
stronger, severely depleting the hot phase.  Pressure and densities
are low, collapsing clouds are very massive (up to $10^6$ \msun) and
\poro\ is high (even higher than unity).  The hot phase is so strongly
depleted that eventually, after roughly one infall time, \fh$<$\fc.
When the cold phase percolates the volume, the most likely outcome is
collapse (the total mass of cold gas is surely larger than $m_{\rm
J}$) and a sudden burst of star formation without any obvious external
trigger.  In other words, star formation would switch from a
``candle''-like to a ``bomb''-like solution.  For a dynamical time
\tff\circa $5\times 10^7$ yr (in this case \nc\circa 1 \cmt) and for
an efficiency \fstar\circa0.1, $10^{10}$ \msun\ of cold gas would give
rise to a starburst of tens of \msunyr.

Another example of critical behaviour is found when a high-density
system goes into the PDS confinement regime (see the previous
subsection).  If the mixing of hot phase and hot internal gas is fast,
then more thermal energy is lost to radiation and \fpds=$0.737(1
-(\Tfin/\Tpds)^{-3.2}) +0.263$ (Equation~\ref{eq:defbpds}).  In this
case, when PDS confinement is triggered \th\ decreases dramatically,
and the hot phase collapses in a very short time.
Fig.~\ref{fig:pdsgno} shows the evolution of the same example of
Fig.~\ref{fig:pdscon} for this choice of \fpds.  The collapse of the
hot phase takes place at the start of PDS, after 1.5 infall times (1.5
Gyr). The cold gas will be promptly consumed into stars; for a
conservative value of \fstar=0.1 this will give rise to a brief star
formation episode with \dmsfr\ in excess of 100 \msunyr.  The porosity
of SBs takes values larger than unity at PDS, indicating the formation
of a unique super-SB that will plausibly remove all the ISM not
consumed by star formation.  This super-wind will interact with the
external halo gas, so that further infall will be halted for some
time.

The actuality of these critical solutions is uncertain, as they could
be due to some of the simplifications introduced.  In low-density
cases, magnetic fields or turbulence could keep the cold medium
fragmented even in the presence of low thermal pressure, or the hot
phase could be replenished by mass flows neglected here, while in
high-density cases the existence of critical solutions depends
sensitively on the way the energy of SNe exploding after the SB stalls
is given to the ISM.  In any case, the idea of a critical ISM deserves
further investigations.

\subsection{Probing the parameter space}

In the following we give a brief account of the effects of changing
the various parameters within reasonable limits.

The value of \sigv\ enters formally in the coagulation time
(Equation~\ref{eq:coag1}), so a hypothetical increase of \sigv\ at
fixed \heff\ and \stot\ reflects in a decrease of the coagulation
time, and an increase of the largest collapsing mass $m_{\rm u}$
(Equation~\ref{eq:upper}), with a corresponding increase of the
collapsing mass \mcc\ (Equation~\ref{eq:mcc}), the fraction of cold
gas available for collapse \fcoll\ (Equation~\ref{eq:fcoll}) and a
resulting higher (and more quickly decreasing) star formation rate
(Equation~\ref{eq:dmsfr}).  However, the collapsing mass \mcc\ is set
mostly by the Jeans mass, so the results are rather insensitive to the
precise value of \sigv.  Of course, in real systems a change in \sigv\
would imply a change in \heff\ or \stot, with the known effects.

Decreasing \tinf\ results in a correspondingly stronger star-formation
rate, and in a faster recycle of materials, while increasing it has an
opposite effect.  Again, the feedback regimes do not change much; with
fast infall adiabatic confinement is reached at slightly lower
densities.

The effect of decreasing \alfa\ is that of moving mass to the
high-mass end of the mass function of cloud, and thus to increase
\fcoll\ and \dmsfr.  An increase of the star-formation rate is
obtained also by increasing \fstar\ (Equation~\ref{eq:dmsfr}).
However, the two cases are rather different in terms of cold gas: with
low \fstar\ and \alfa\ values the cold gas is reprocessed by collapse
many times, while with high values it is locked in the small clouds
until it is processed by star formation.  On the feedback regimes,
lowering \alfa\ or \fstar\ has the effect of moving the limit for
adiabatic blow-out at lower densities, especially at low \heff\
values, and vice-versa for an increase of \alfa\ or \fstar.

An increase of \shape\ to unity lowers the Jeans mass
(Equation~\ref{eq:lombert}), making blow-out more difficult; also, a
lower number of SNe per cloud in the adiabatic confinement case is
obtained.  The opposite happens for a decrease of \shape, which is
justified if collapsing clouds are supported by kinetic or magnetic
pressure; for \shape=0.05 adiabatic blow-out is easily reached at
densities higher by and order of magnitude than the limit given in
Equation~\ref{eq:limit_bo}.

As mentioned above, a low value of \esn\ leads to lower \th, lower
pressure and a corresponding increase of the Jeans mass; moreover, PDS
confinement is found at high \heff\ and \stot\ values.  Increasing \esn\
leads to an increase of \th, that for \esn=10 can reach extremely high
and unlikely values.  The limit between the adiabatic blow-out and
confinement regimes does not depend much on \esn.

As cooling is a relatively modest mass flux, the results do not depend
sensitively on the parameter \fcool, with the exception that PDS
confinement is reached more easily for \esn=1 whenever \fcool$\ll$1.

The parameters \fevap\ and \tevap\ mostly affect the adiabatic
blow-out regime; in particular \tevap\ influences strongly the
resulting \th.  A low value of \fevap\ increases the number of
critical blow-out cases, until \tevap\ is increased to compensate for
the lack of evaporated mass.  If both parameters are increased, the
limit for adiabatic blow-out lowers.  Finally, with a high \fevap\
value PDS confinement is reached at high densities even for \esn=1.

\subsection{On the vertical scale-height of the hot phase}

The assumption of one single vertical scale-height for both cold and
hot phases is clearly over-simplistic; it can be relaxed by assuming
two different scale-heights, but in absence of further constraints
this new degree of freedom would not contribute significantly to the
understanding of the problem.  Anyway, the assumption is sensible as
the hot gas is continually replenished within \heff.  The hot gas that
leaks out into the halo is likely to settle in a low-density layer
that surrounds the galaxy.  Such layer is observed in the Milky Way
(see, e.g., Jenkins 2002) as well as in nearby galaxies (see, e.g.,
Ferguson et al. 1995; Fraternali et al. 2002).  The presence of a
sufficiently steep decreasing density gradient at \heff\ is enough to
guarantee that this layer does not hamper the blow-out of SBs.
Indeed, the interaction of blown-out gas and such a layer could be at
the origin of the observed correlation of X-ray and H$\alpha$ fluxes
in nearby starburst galaxies (Strickland et al. 2002).

We can estimate the thickness of this layer, that we call \hfin, as
follows.  We assume that the expansion of the leaked-out gas stops
after one cooling time.  For a gas in adiabatic expansion we have
$\Nh'= \Nh\Heff/\Hfin$, $\Th' = \Th (\Heff/\Hfin)^{\gamma-1}$ and
$c_{\rm s,h}'= c_{\rm s,h} (\Heff/\Hfin)^{(\gamma-1)/2}$, where the
prime indicates the quantities relative to the gas layer and $\gamma$
is the adiabatic index, assumed to be 5/3.  It is easy to see that,
for the cooling function given in Equation~\ref{eq:coolfunc} the
cooling time is constant with \hfin.  The equilibrium value of \hfin\
will satisfy the condition $\Tcool = \Hfin/c_{\rm s,h}'$.  To compute
it, we take into account that there is continuous injection of hot gas
with a roughly constant rate within one cooling time.  We obtain:

\be \Hfin = \frac{4}{7} \Heff \left( \frac{\Tcool c_{\rm s,h}}{\Heff}
\right)^{3/4} \, . \label{eq:hfin} \ee

\noindent
For the adiabatic blow-out case of Fig.~\ref{fig:stadbo} we obtain
\hfin\circa2 \kpc.  This will be an overestimate, as gravity is likely
to be important at such distances from the centre of the galaxy.

Alternatively the expansion of the hot phase could progress in such a
way not to create a steep density gradient.  In this case the hot
phase would be contained in a layer of thickness \hfin\ and nearly
constant density.  Exploiting Equation~\ref{eq:hfin}, it is possible
to include the dynamical evolution of such a layer in the system of
equations; in this case leak-out would not be considered as the hot
phase formally never gets back to the halo.  As a result, the hot gas
produced by SNe pushes the layer to high enough \hfin\ values that
adiabatic confinement is always achieved.  The predictions of this
version of the model are found in striking disagreement with
observations in the Milky Way-like case (see next section), so the
possibility of a hot gas layer with a roughly constant density will
not be further considered.

\section{Discussion}

\subsection{The Milky Way}

To highlight the predictive power of the model, we compare the results
of the Milky Way-like adiabatic blow-out case shown in
Fig.~\ref{fig:stadbo} with available observational evidence.  It is
useful to stress that no accurate modeling or fine-tuning of
parameters is attempted at this stage, so an order-of-magnitude
agreement is considered a success.  The predictions of \nh\
(\circa$10^{-3}$ \cmt), \nc\ (\circa10 \cmt) and \th\ (\circa\
$3\times 10^6$ \K) are in broad agreement with the ISM of the Milky
Way.  Thermal pressure (\circa$10^3$ \K\ \cmt) is in line with
observations, but is an order of magnitude lower than the observed
total pressure, which is dominated by turbulent and magnetic
contributions.  The mass ratio of cold gas to stars (\circa0.1) is
correctly reproduced after 10 \Gyr.

The star formation rate \dmsfr\ is slowly decreasing in time, with a
ratio of average to final rates of \circa3.  This roughly consistent
with the results of the chemical evolution model of Chiappini et
al. (1997).

Both the average value and the range of the masses of collapsing
clouds are smaller than those observed for molecular clouds, that can
be as large as $m\cl\ga10^6\ {\rm M}_\odot$.  Large collapsing clouds are
easily obtained by using a very low value for \shape, on the ground
that kinetic support determines the Jeans mass $m_{\rm J}$ (Section
2.3).  The small range of collapsing cloud masses reflects in the low
values of \fcoll, the fraction of cold mass in collapsing clouds; it
is predicted to be \circa5 per cent, at variance to the observed
\circa50 per cent.  To obtain higher \fcoll\ values it is useful to
decrease \alfa\ to the observed value of 1.6, decreasing also \fstar\
to avoid excessive star formation (this is also consistent with
observations).  However, good \fcoll\ values are obtained only by
allowing clouds to coagulate for at least 10 dynamical times.  This
unrealistic value is not worrisome if we consider the uncertainty
connected to the coagulation picture; while the coagulation time,
\tcg\circa$5\times 10^7$ yr, is coincidentally similar to the time
interval between the sweeping of two spiral arms, the formation of
molecular clouds in the converging flows of spiral arms
(Ballesteros-Paredes, Vazquez-Semadeni \& Scalo 1999) could easily be
more efficient than random aggregations of clouds, and this could be
mimicked by allowing coagulation to work for many dynamical times.

The thickness of the layer of leaked-out hot gas is predicted to be
\circa2 \kpc, in rough agreement with the value of \circa3 \kpc\
estimated for the Milky way by Savage et al. (2000; see also Jenkins
2002).  However, this estimate is based on FUSE detection of O VI
absorption lines of OB stars; this method is sensitive to temperatures
in a narrow range around \circa$3\times 10^5$ \K.  The gas leaking out
at \th=$10^6$ \K\ has adiabatically cooled to \circa$1.5\times10^5$
\K\ at 2 \kpc, while the temperature of \circa$3\times 10^5$ \K\ is
reached at \circa0.6 \kpc, significantly less than observed.  However,
this prediction depends sensitively on the parameter \tevap, that
influences the temperature of the hot phase.  If this parameter is
increased by a factor of 3 (a reasonable choice according to paper
II), the resulting layer is predicted to be 10 \kpc\ thick, reaching a
temperature of \circa$3\times 10^5$ \K\ at \circa3 \kpc\ as observed.

With reasonable choices of the parameters, and allowing coagulation to
work for 10 dynamical times, it is possible to reproduce all these
properties of the Milky Way.  The reason why we do not stress this
result is because we consider the present model too simple to draw
significant conclusions from it.  By interfacing this model with an
algorithm for disc formation in a cosmological dark-matter halo and
including the effect of differential rotation and spiral arms it will
be possible to produce accurate predictions for the Milky Way,
including galactic fountains, high-velocity clouds, chemical
enrichment of the various components, chemical gradients along the
disc and so on.  By reproducing the observed Milky Way it will be
possible to constrain most model parameters by modeling just one
galaxy.

\subsection{Critical solutions and the triggering of galactic winds}

Although a proper modeling of galactic winds requires specifying the
properties of the dark-matter halo hosting the galaxy, it is
interesting to analyse the cases in which feedback could lead to the
removal of a significant quantity of ISM from a galaxy.  As removal of
gas from a halo with low circular velocity can be achieved even with a
single SB (see, e.g., Ferrara \& Tolstoy 2000), we will concentrate on
bright galaxies.  Blow-out leads to the expulsion of matter with a
velocity that, in the example of Fig.~\ref{fig:stadbo}, ranges from
\circa250 \kms\ at 1 Gyr to more than 500 \kms\ at later times, so if
these clouds are not slowed down significantly by the halo gas
(e.g. by the layer of leaked-out hot gas) they may escape even from
relatively high-mass halos.  Anyway, blow-out flows are never very
strong, so blow-out is unlikely to lead to massive removal of ISM from
a galaxy.  Besides, leaked-out gas cools below 10$^5$ \K\ at \circa10
kpc, so it will be emitted as a tenuous wind from the low-mass halos,
but will be retained by the halo of a bright galaxy.  Leaked-out gas
is much hotter in the adiabatic confinement regime, so this gas will
be able to escape from moderate-sized halos, but will be retained for
instance in big elliptical galaxies.  In conclusion, as long as blasts
propagate into the hot phase and SBs do not percolate the volume, the
removal of mass is inefficient in bright galaxies.

This conclusion changes in the critical cases, where a significant
amount of gas accumulated for some time is consumed in a few dynamical
times, and when the porosity of SBs gets unity value.

Critical solutions are found at least in three cases: (i) thin, very
low-\stot\ systems in the adiabatic blow-out regime, (ii) thick,
moderate-\stot\ systems in the adiabatic blow-out regime (in case of
very efficient blow-out) and (iii) thick, high-\stot\ systems in the
PDS confinement regime (with low \esn\ and low \fpds).  While case (i)
has no astrophysical relevance (such systems would anyway be kept
ionized by the cosmological UV background), case (ii) may correspond
to some gas-rich dwarf galaxies and case (iii) to high-redshift
spheroids.

When these systems become critical, all the cold phase collapses and
gives rise to diffuse star formation.  For a conservative \fstar\
value of 0.1, we estimated star-formation rates of tens of \msunyr\
for case (ii) and in excess of 100 \msunyr\ for case (iii).  In such
big bursts \fstar\ could well take unit values, thus boosting star
formation rates even higher; on the other hand, if the transition from
the ``candle''- to the ``bomb''-like regime is not as quick as
assumed, star formation rates will be lower.  Analogously to what
happens in star-forming clouds, the exploding SNe will propagate into
the diffuse cold phase, going soon in the PDS stage and then
percolating the volume.  This will give raise to a unique SB with a
very high mechanical luminosity, able to sweep the whole galaxy.  This
snowplow will eventually blow out of the galaxy and then fragment
because of Raileigh-Taylor instabilities.  If the momentum of the gas
in the fragmented snowplow at this point is sufficient, it will be
thrown out of the galaxy.

Percolation of SBs gives a similar effect if it takes place in the PDS
confinement (or blow-out) regime; though it has been assumed for
simplicity that clouds pierce the snowplow, this is likely true only
for the largest and densest clouds, that would likely be star-forming
in this case.  The effect of a percolation of collapsed shells would
be to create a super-SB that sweeps the ISM, pushing part of the gas
out of the galaxy in form of cold clouds while the rest is compressed
toward the centre of the galaxy.  This is found, for instance, in the
simulations of primordial galaxies by Mori, Ferrara \& Madau (2002);
as the physics is the same, their conclusion can be extended to
larger, lower-density galaxies, as long as percolation of SBs in the
PDS confinement regime is obtained.  Obviously, the gas concentrated
at the centre would give rise to a secondary burst of star formation
that would pump further energy into the super-SB.

Percolation of SBs in the adiabatic stage is likely to have a smaller
effect, as the blast would continue to propagate into the hot phase.
The cold phase would be affected by the relatively inefficient
processes of thermo-evaporation and cloud dragging (preferentially in
the radial direction).  These same processes are in place also in
presence of a hot phase that continually leaks out of the galaxy.  In
the adiabatic confinement regime the cold phase is so dense and with
such a low filling factor that these effect are likely to be
inefficient, while in the case of adiabatic blow-out from a disc the
dimensionality of the system would presumably lead to a funneling of
the energy in the vertical direction, thus making a massive removal of
gas unlikely.

As a matter of fact, the condition \poro$>$1 is never met in the
adiabatic confinement regime, while it is achieved in the critical
cases discussed above (Fig.~\ref{fig:pdsgno}).

There are other cases in which the system may become critical.  In
presence of very high pressure the density of the cold phase can be so
high (say $>$10$^3$ \cmt) that the formation of H$_2$ is triggered
even in absence of collapse.  If this limit is reached when much gas
is accumulated, this may give raise to a sudden burst of star
formation.

Finally, a critical behaviour of the system can be triggered from
outside.  For instance, a strong tidal perturbation (or a merger) would
act in two important ways, by lowering the Jeans mass of the clouds
(because of the external pressure) and by thickening the structure,
thus allowing a disc-like system to switch from adiabatic blow-out to
adiabatic confinement; this would lower the Jeans mass even more and
decrease the dynamical time.  The effect would be a rapid consumption
of the accumulated cold phase and a likely percolation of SBs.

\subsection{Simplified models}

The present model can be generalized to reproduce the components of
real galaxies, like disc, bulge and halo, and then interfaced with a
galaxy formation code that includes the mass assembly of dark-matter
halos, cooling inside those halos, disc formation, galaxy mergers,
interaction with galaxy clusters etc..  However, it is much more
convenient to device a set of approximate analytic solutions to this
feedback model.  These solutions can also be adapted to model the
``sub-grid'' physics of feedback in N-body simulations.

The solutions in the \heff--\stot\ plane can be divided into four main
regions where different regimes are met (adiabatic blow-out, adiabatic
confinement, PDS confinement, critical blow-out cases).  These regions
are separated by limiting relations of the kind $\Stot = \Sigma_{\rm
tot,0} (\Heff/1\ {\rm kpc})^{-\alpha_{\rm lim}}$, where the exponent
${\alpha_{\rm lim}}$ is usually in the range 0.5--1.  At the lowest
densities (a factor 10$^2$ lower than Equation~\ref{eq:limit_bo} for the
reference choice of parameters) systems are critical, but they will
most likely be kept ionized by the cosmological UV background, so they
will simply not evolve.  At densities below Equation~\ref{eq:limit_bo}
(again for the reference choice of parameters) the system is in the
adiabatic blow-out regime.  Above that limit it gets into the
adiabatic confinement regime.  PDS confinement is reached at densities
higher than Equation~\ref{eq:limit_pds} (valid for \esn=0.3 and \fevap=0.05).

For each regime a simplified solution can be obtained by noticing the
following facts that are found to hold in most cases: 1) \th\ is
nearly constant, and equal to a value that mostly depends on the
regime and on \esn; 2) $F_{\rm h}$, \fcoll\ and \fc\ are nearly
constant to values that mostly depend on the regime; 3) cooling is
negligible in all cases but those in PDS confinement, where leak-out
is negligible; 4) leak-out dominates over blow-out in non-critical
cases if \fbo\ is not large.  With these assumption it is relatively
easy to solve the system of Equations~\ref{eq:massfluxes} for the mass
flows, while typical values of \th, $F_{\rm h}$, \fcoll\ and \fc\ the
different feedback regimes have been given in Section 4.  However, a
proper presentation of these simplified solutions requires some
discussion that is out of place in this paper, so they will be
presented elsewhere.

\subsection{Limitations and further work}

The merit of such relatively simple modeling is to highlight the
possible physical regimes one should expect once a more complete
calculation is performed.  However, there are a number of limitations
that have to be carefully taken into account to assess the validity of
the results presented here.

(i) The Sedov solution for the SBs in the adiabatic stage is only a
rough approximation of reality.  There is a long list of effects,
mentioned above and in part described by Ostriker and McKee (1988),
that influence the dynamics of SBs.  However, as long as the SBs
expand in the relatively smoother hot component, it is likely that the
Sedov solution gives the roughly correct evolution and functional
dependences for the SBs.

(ii) As already mentioned, thermal conduction at the interface of cold
and hot phases can make part of the cold gas evaporate.  It has been
verified that the impact of thermo-evaporation is small in the mass
flows even if it is not quenched by magnetic fields, as the
thermo-evaporated mass is usually smaller than the evaporated mass of
the collapsing cloud whenever \fevap\ is not much smaller than one.

(iii) Type Ia SNe have not been considered.  However, their
introduction is straightforward in this model; they will interact
directly with the ISM through a set of uncorrelated SNRs.  As shown by
Recchi et al. (2002), Type Ia SNe may be very important because they
explode {\it after} a burst of star formation, and can contribute to
maintain the hot phase when most cold gas is consumed.

(iv) Cosmic rays are known to be in rough equipartition with
turbulence and magnetic fields.  They are accelerated by the shocks
generated by SNRs and SBs, directly or indirectly through turbulence
(see, e.g., Longair, 1981).  The role of cosmic rays, which are
confined within the galaxy by magnetic fields, is that they distribute
their energy to all the ISM, and not only to the densest collapsing
clouds.  So, they could give an important contribution to the mass and
energy flows.

(v) There are other channels of mass and energy exchange between
components that we are not considering here.  One is the decay of
turbulence driven by the kinetic energy of SNe, the other is the
presence of a significant amount of mass in a warm phase, that can
receive part of the energy of the blast and radiate it.  Also, UV
light coming from massive stars or from an external UV background
could be responsible (together with thermo-evaporation) for continuous
evaporation of the cold phase.  Although some analytical estimates of
these effects are possible, accurate numerical simulations will be
necessary to assess the importance of these processes.

(vi) In realistic situations the ISM is subject to many influences,
like spiral arms, differential rotation, tidal disturbances, mergers,
ram pressure from hot halo gas (like in ellipticals or clusters) etc..
All these processes can be modeled once the global structure of the
galaxy and its environment are specified.  For instance, the passage
of a spiral arm can be modeled by a periodic decrease of the Jeans
mass due to the external pressure term.  As in the adiabatic blow-out
case the coagulation time is \tcg\circa$5\times 10^7$ \yr, of the
same order of the frequency of spiral arms, the clouds have just time
to coagulate to increase their mass by a factor of a few before the
spiral arm sweeps again.  So, a moderate decrease of the Jeans mass
would suffice in guaranteeing that star formation takes place mainly
in the spiral arms, even in absence of a more explicit connection,
like that proposed by Ballesteros-Paredes et al. (1999).

(vii) The model presented here is assumed to be valid in the regime
where many generations of collapsing clouds self-regulate in forming a
galaxy.  For dwarf galaxies some changes in the model are necessary:
firstly it is important to check that at least one collapsing cloud is
present; secondly, the first episode of star formation could itself
cause a complete blow-away of the ISM (see, e.g., Ferrara \& Tolstoy
2000), so that the system may never get into a self-regulated regime.
Moreover, for dark matter halos with small circular velocities the
leaked-out or blown-out gas will most likely be lost to the
inter-galactic medium.

(viii) This model gives satisfactory predictions for the state of the
ISM in a Milky Way-like situation.  The high-density, bursting cases
are subject to much weaker observational constraints, not only for the
paucity of very nearby starbursts but also for the presence of dust
that hampers observations in the optical, UV and soft-X bands.
Besides, the extrapolation of the assumptions that are successful at
low densities is not straightforward.  For instance, large collapsing
clouds could be generated in cooling flows or, as mentioned above,
star formation could be triggered even in clouds smaller than the
Jeans mass when thermal pressure makes them denser than a threshold
density at which H$_2$ starts to form.  Careful comparison with
available observations is needed to constrain the parameters of
feedback in the starburst cases.

(ix) This model can be improved to give more accurate predictions on
observables related to the ISM.  This would require proper (numerical)
modeling of the intermediate warm phase(s) and of the ionization
equilibrium between the phases.

\subsection{Comparison with previous works}

The model of the ISM of McKee \& Ostriker (1977) has been a reference
model for years, although the picture based on compressible supersonic
MHD turbulence is now emerging.  The model presented here has many
points in common with McKee \& Ostriker (1977), but presents many
improvements: (i) we address the dynamics of the ISM, including star
formation and feedback; (ii) we assume no equilibrium, but investigate
on the conditions that lead to self-regulated or critical ISM; (iii)
we take into account the correlated nature of Type II SNe.  Besides,
we do not consider the warm phase and its ionization equilibrium with
the hot phase.  In the present model we do not require unit porosity
of SBs to justify its presence of a hot phase (which sometimes cannot
even be maintained).  As a matter of fact, unit porosity of SBs is
hardly reached in non-critical cases; however, uncorrelated adiabatic
SNRs that stop at \rbos\ would have a porosity of order one in the MW
example, but not in the EL one in the adiabatic confinement case.  So,
on the light of the present results, $Q=1$ is at best an unnecessary
assumption.

Some recent works on galaxy formation by Silk (1997; 2000), Efstathiou
(2000), Shu, Mo \& Mao (2003) or Springer \& Hernquist (2003) present
models of feedback and star formation based at least in part on the
McKee \& Ostriker (1977) model.  In these cases Type II SNe are
assumed to be uncorrelated and the ISM is assumed to be self-regulated
to a unit value of the porosity of SNRs.  For instance, Efstathiou
(2000) fixes the star formation by assuming equilibrium between the
kinetic energy acquired by cold clouds at shocks and that lost by
coagulation, while Silk (1997, 2000) connects star formation to the
dynamics of the disc by requiring the Toomre $Q$-parameter to be unity
and postulating the identity of the time scale for star formation with
the viscous time scale; this is an important ingredient for obtaining
exponential discs, but the nature of this identity remains
unexplained.  The model presented here does not assume any
equilibrium, and does not use any ingredient of disc dynamics, thus
being applicable in virtually all situations.  While it is clear that
disc dynamics will influence the evolution of a spiral galaxy, our
results suggest that most properties of galaxy formation can be
understood simply as a chain of local processes.

An alternative to the present modeling is to consider the ISM as
turbulent.  As shown by Avila-Reese \& Vazquez-Semadeni (2001) the ISM
can be considered as a globally turbulent medium, with turbulence
forced in specific places (the star-forming regions) and propagating
throughout the volume.  The ``diffusion'' velocity of turbulence is
connected to the time scale of decay of turbulence.  Both this group,
that uses a 2D code, and Mac Low et al. (1998), who use a 3D code,
find that turbulence decays as $t^{-\alpha}$ with $\alpha\simeq0.8$.
It is easy to show that (Avila-Reese \& Vazquez-Semadeni 2001), when
turbulence is forced in some specific sites, the rms velocity of
turbulence scales with distance from the forcing region as
$u_{rms}\propto l^{-\alpha/(2-\alpha)}$, while the decay distance of
turbulence grows with time as $l\propto t^{1-\alpha/2}$.  It is
remarkable that for $\alpha=0.8$ the two exponents are exactly equal
to our relations $R_{\rm sb}\propto t^{0.6}$ and $v_{\rm sb}\propto
R^{-0.4}$.  While a direct physical interpretation of this fact may be
misleading without further investigation, it is clear that the
propagation of energy through the ISM by isolated spherical blasts is
not in clear contradiction with the results of the turbulent model.
This confirms the validity of a simple treatment as a first
approximation.

\section{Conclusions}

We have presented a model for feedback in galaxy formation, based on a
two-phase ISM, that does not restrict to self-regulated, equilibrium
solutions and neglects (for simplicity) the global structure of the
galaxy, apart from its density, vertical scale-height and velocity
dispersion of clouds.  From the dynamics of the SBs that arise from
the collapsing ``molecular'' clouds, we have identified four possible
regimes of feedback, depending on whether SBs blow out of the ``disc''
or remain pressure-confined, and whether they have time to enter the
PDS stage.  For a reference set of parameter values we have studied
the dynamics of the system in the vertical scale-height -- surface
density plane, identifying the regions of the plane corresponding to
different regimes.  Both blow-out and confinement mostly take place in
the adiabatic regime.  In a Milky Way-like adiabatic blow-out case,
the main characteristics of the ISM of the Galaxy are broadly
reproduced.  In the adiabatic confinement regime the ISM is predicted
to have higher pressure, temperature of hot phase and densities of
both phases, and smaller collapsing clouds; in some cases the density
of the cold phase could be high enough to trigger diffuse star
formation.  PDS confinement is found for high-density, thick
structures in significant regions of the parameter space.  In this
case feedback is less effective, the hot phase cooler and star
formation quicker.

In many cases the system becomes critical, in the sense that the hot
phase is severely depleted and the cold phase percolates the whole
volume.  This happens for very low-density thin systems (that would
however be kept ionized by the cosmological UV background), in some
regions of the parameter space also for low-density thick systems in
adiabatic blow-out (that may correspond to some gas-rich dwarf
galaxies) and for high-density thick systems in PDS confinement (that
may correspond to high-redshift galaxies).  The most likely result of
this critical behaviour is the sudden consumption by star formation of
the cold gas accumulated by the galaxy; the dynamics switches from a
``candle''- to a ``bomb''-like solution.

The porosity of SBs is usually found to be much lower than unity.
However, in some cases unit porosity is found while SBs are in the PDS
stage.  This corresponds to the formation of a super-SB that sweeps
the whole galaxy, removing most ISM from it.  These events, together
with the critical solutions, are likely connected to the triggering of
galactic winds.

With respect to previous models of feedback, the main parameters that
are typically present, as the efficiency of feedback, the Schmidt law
with its normalization, or the rate of blow-out and leak-out of gas
from a star-forming galaxy, are predictions of the present model.  The
parameter space is connected to the properties of the ISM, and can
thus be constrained by observations of the Milky Way and nearby
galaxies; most parameters can be fixed in principle by reproducing
only the Milky Way.  Moreover, the mass flows used in this model can
be fine tuned by comparing with future detailed simulations of the ISM
in a forming galaxy that include all the main physical processes
though to be at work.

This model does not restrict to self-regulated ISM, and presents a
rich variety of solutions with a relatively limited set of parameters.
Although the turbulent nature of the ISM is not explicitly taken into
account, the model is thought to give a good approximation to the
solution of the feedback problem.  The feedback regimes found here can
be used, together with the refinements of the model that will be given
in upcoming papers, to construct a realistic grid of feedback
solutions to be used in galaxy formation codes, either semi-analytic
or numeric.

\section*{Acknowledgments}

The author thanks Andrea Ferrara, Gianrossano Giannini, Gabriele
Cescutti and especially Simore Recchi for many fruitful discussions.

{}


\appendix
\section[]{List of frequently used symbols}

\begin{tabular}{@{}ll}

\alfa          & Slope of mass function of clouds          \\
$a\cl$         & Radius of cloud                           \\
$c_{\rm s,h}$  & Sound speed of the hot phase              \\
\esn           & Energy released by a SN                   \\
\eth           & Thermal energy of a SB                    \\
\ekin          & Kinetic energy of a SB                    \\
\desn          & Rate of energy release from SNe           \\
\decool        & Rate of energy loss by cooling            \\
\desnpl        & Rate of energy loss by snowplows          \\
\debo          & Rate of energy loss by blow-out           \\
\deleak        & Rate of energy loss by leak-out           \\
\desb          & Rate of energy gain from SBs              \\
\dehot         & Net energy flux of the hot phase          \\
\fe            & Efficiency of feedback                    \\
\fh,\fc        & Filling factors of the two phases         \\
\fcoll         & Fraction of cold gas in collapsing clouds \\
\fstar         & Efficiency of star formation              \\
\fevap         & Evaporated fraction of collapsed cloud    \\
\fbo           & Fraction of swept gas blown out by a SB   \\
\fbom          & Largest value of \fbo                     \\
\fcool         & Fraction of cooled gas in a cooling flow  \\
\frest         & Fraction of restored mass                 \\
\fpds          & Release of energy after PDS confinement   \\
$F_{\rm h}$    & Fraction of hot gas                       \\
\heff          & Vertical scale-height of the system       \\
\heffh         & Dynamical vertical scale-height of hot gas\\
\hfin          & Height of the layer of hot leaked-out gas \\
\lmech         & Mechanical luminosity of a SB             \\
$m\cl$         & Mass of clouds                            \\
$m_{\rm l}$    & Lower cutoff mass of clouds               \\
$m_{\rm u}$    & Upper cutoff mass of clouds               \\
$m_{\rm J}$    & Jeans mass of clouds                      \\
\mcc           & Mass of the collapsing cloud              \\
$M_{\rm sw}$   & Mass swept by a SB                        \\
$M_{\rm int}$  & Swept mass that is still hot inside a SB  \\
\mstsn         & Mass of formed stars per SN               \\
\mtot          & Total mass of the system                  \\
\uno           & Mass of the i component$^\ast$            \\
\due           & Mass of metals in the i component$^\ast$  \\
\tre           & Net mass flux of the i component$^\ast$   \\
\qua           & Net metal flux of the i component$^\ast$  \\
\end{tabular}

\noindent
\begin{tabular}{@{}ll}

\dminf         & Infall rate                               \\
\dmcool        & Cooling rate                              \\
\dmleak        & Leak-out rate                             \\
\dmsfr         & Star formation rate                       \\
\dmrest        & Restoration rate                          \\
\dmev          & Evaporation rate                          \\
\dmint         & Sweeping rate minus snowplow rate         \\
\dmsnpl        & Snowplow rate                             \\
\muh,\muc      & Molecular weights of the two phases       \\
\shape         & Shape parameter for collapsing clouds     \\
\nh,\nc        & Density of the two phases                 \\
$N\cl$         & Mass function of clouds                   \\
\ncc           & Total number of collapsing clouds         \\
\nsn           & Number of SNe in a collapsing cloud       \\
$P$            & Pressure of the ISM                       \\
\poro          & Porosity of SBs                           \\
$R_{\rm sb}$   & Radius of SB                              \\
\rpds          & Radius of shell collapse for a SB         \\
$R_{\rm conf}$ & Confinement radius of a SB                \\
\rbos           & Final radius of a blown-out unconfined SB  \\
\rfin           & Final radius of a SB                       \\
\rsn            & Rate of SN explosions in a collapsing cloud\\
\ro             & Total density of the system               \\
\roh,\roc       & Average densities of the two phases        \\
\sigv           & Velocity dispersion of clouds              \\
\stot           & Total surface density of the system        \\
\tcg            & Coagulation time                           \\
\tff            & Dynamical time of clouds                   \\
\tcool          & Cooling time of hot gas                    \\
\tcross         & Sound crossing-time of a SB                \\
\tpds           & Time of shell collapse for a SB            \\
$t_{\rm conf}$  & Confinement time for a SB                  \\
\tlife          & Lifetime of an 8 \msun\ star               \\
\tbo            & Time of first blow-out of a SB             \\
\tbos           & Final time of a blown-out unconfined SB    \\
\tfin           & Final time of a SB                         \\
\tinf           & Infall time scale                          \\
\tleak          & Leak-out time scale                        \\
\tsf            & Star-formation time scale                  \\
\th,\tc         & Temperature of the two phases              \\
\tevap          & Temperature of evaporated mass             \\
$v_{\rm sb}$    & Velocity of SB                             \\
$y$             & Yield from massive stars                   \\
$Z_{\rm i}$     & Metallicity of the i component$^\ast$      \\
$\zeta_{\rm m}$ & Metallicity in solar units                 \\

\end{tabular}

\noindent
Notes: \\
$^\ast$ i = hot, cold, $\star$, halo \\

\section[]{Time-scales for the coagulation of cold clouds}

We obtain here the time-scales for coagulation, given in Section 2.4,
from the Smoluchowski equation of kinetic aggregations.  This
demonstration follows Cavaliere et al. (1992).  The Smoluchowski
evolution equation for the mass function of clouds $n(m;t)$ is:

\begin{eqnarray}\lefteqn{\frac{\partial n}{\partial t} = \frac{1}{2} \int_0^m dm' K(m',m-m')
n(m';t)n(m-m';t)} \nonumber \\&& - n(m;t)\int_0^\infty dm' K(m,m') n(m';t)\, .
\label{eq:smolu} \end{eqnarray} 

\noindent
The kernel for aggregations of clouds 1 and 2, $K(m_1,m_2)$ is given
by Equation~\ref{eq:kernel}, where the cross-section for interactions is
given by Equation~\ref{eq:crosssect}.  We define a typical mass for the
mass function $m_\star$ (which is then identified with the upper
cutoff $m_{\rm u}$) and scale all masses with $m_\star$ through the
unidimensional variable $\nu=m/m_\star$.  We separate the kernel for
aggregation of clouds into geometric and resonant terms (the latter
term being considered only in this appendix), and write the two terms
as follows:

\begin{eqnarray}
K_{\rm geom} &=& {\mathcal{F}}_{\rm geom}\, m_\star^{2/3}\,
\langle(\nu_1^{1/3} + \nu_2^{1/3})^2\rangle_{\rm m} \label{eq:kernels} \\
K_{\rm res}  &=& {\mathcal{F}}_{\rm res}\, m_\star^{4/3}\,
\langle(\nu_1 + \nu_2)(\nu_1^{1/3} + \nu_2^{1/3})\rangle_{\rm m}\, . \nonumber
\end{eqnarray}

\noindent 
We call $\lambda$ the exponent of $m_\star$ in the equations.  The two
${\mathcal{F}}$ functions are respectively ${\mathcal{F}}_{\rm
geom}=\bar{\rho}_{\rm c} \pi (4\pi\rho_{\rm c}/3)^{-2/3} \langle
v_{\rm ap} \rangle_{\rm v}$ and ${\mathcal{F}}_{\rm
res}=\bar{\rho}_{\rm c} 2\pi G (4\pi\rho_{\rm c}/3)^{-1/3} \langle
(v_{\rm ap})^{-1} \rangle_{\rm v}$.  Let's assume that the mass
function is expressible as:

\be n(m;t)dm = m_\star^2 \Phi(\nu) d\nu\, . \label{eq:app} \ee

\noindent
This is valid if the slope of the mass function is fixed and if
$m_{\rm l}\ll m_{\rm u}$.
Inserting this {\it anstatz} into the Smoluchowski equation, and
considering only the time-dependent terms, we easily obtain the
equation:

\be \dot{m}_\star = {\mathcal F} m_\star^\lambda\, . \label{eq:app2}
\ee

\noindent
This is valid for the two coagulation modes.  This equation admits the
solution:

\be m_\star(t) = m_{\star0} (1-(\lambda-1){\mathcal F} 
m_{\star0}^{\lambda-1}(t-t_0))^{1/(1-\lambda)}\, . \label{eq:app3} \ee

\noindent
We can then define a coagulation time as $t_{\rm coag} =1/{\mathcal F}
m_{\star0}^{\lambda-1}$.  For the two coagulation modes we obtain:

\begin{eqnarray} 
m_\star(t) &=& m_{\star0} \left(1+\frac{t-t_0}{3t_{\rm
coag}} \right)^3 \ \ \ \ \ {\rm (GEOM.)} \label{eq:app4} \\
m_\star(t) &=& m_{\star0} \left(1-\frac{t-t_0}{3t_{\rm
coag}} \right)^{-3} \ \ \ \ \ {\rm (RES.)}\, . \nonumber
\end{eqnarray}

\noindent
It is easy to verify that the coagulation time is given by
Equation~\ref{eq:coag1} in the case of geometrical interactions, while
in the resonant case:

\be t_{\rm coag} = \left(\frac{4\pi}{3}\right)^{1/3} \frac{1}{2\pi G}
\bar{\rho}_{\rm c}^{-2/3} \frac{\rho_{\rm c}}{\bar{\rho}_{\rm
c}}^{1/3} \frac{1}{m_{\rm J}^{1/3}\langle (v_{\rm ap})^{-1}
\rangle
}\, . \ee

Finally, it is worth noticing that the slope of the mass function is
assumed not to change during the evolution of the system, while
coagulation would in general imply a flattening of the mass function.
This is reasonable in cases where coagulation and collapse regulate
the mass function to a given shape.

\section[]{Feedback regimes in the \nh--\lmech\ plane}

The fate of the SB depends mostly on the hot phase density \nh\ and
the mechanical luminosity \lmech.  To better quantify the fate of the
SB we restrict to the case of solar metallicity and $\Muh=0.6$, and
express the hot phase density in units of $n_{\rm h,-3}= \Nh/(10^{-3}\
cm^{-3})$.  Moreover, \heff\ is expressed in units of $10^3$ \pc\ and
the temperature of the hot phase in units of $10^6\ K$.  Table C1
reports the characteristic times of SBs.

\begin{figure}
\centerline{
\includegraphics[width=9cm]{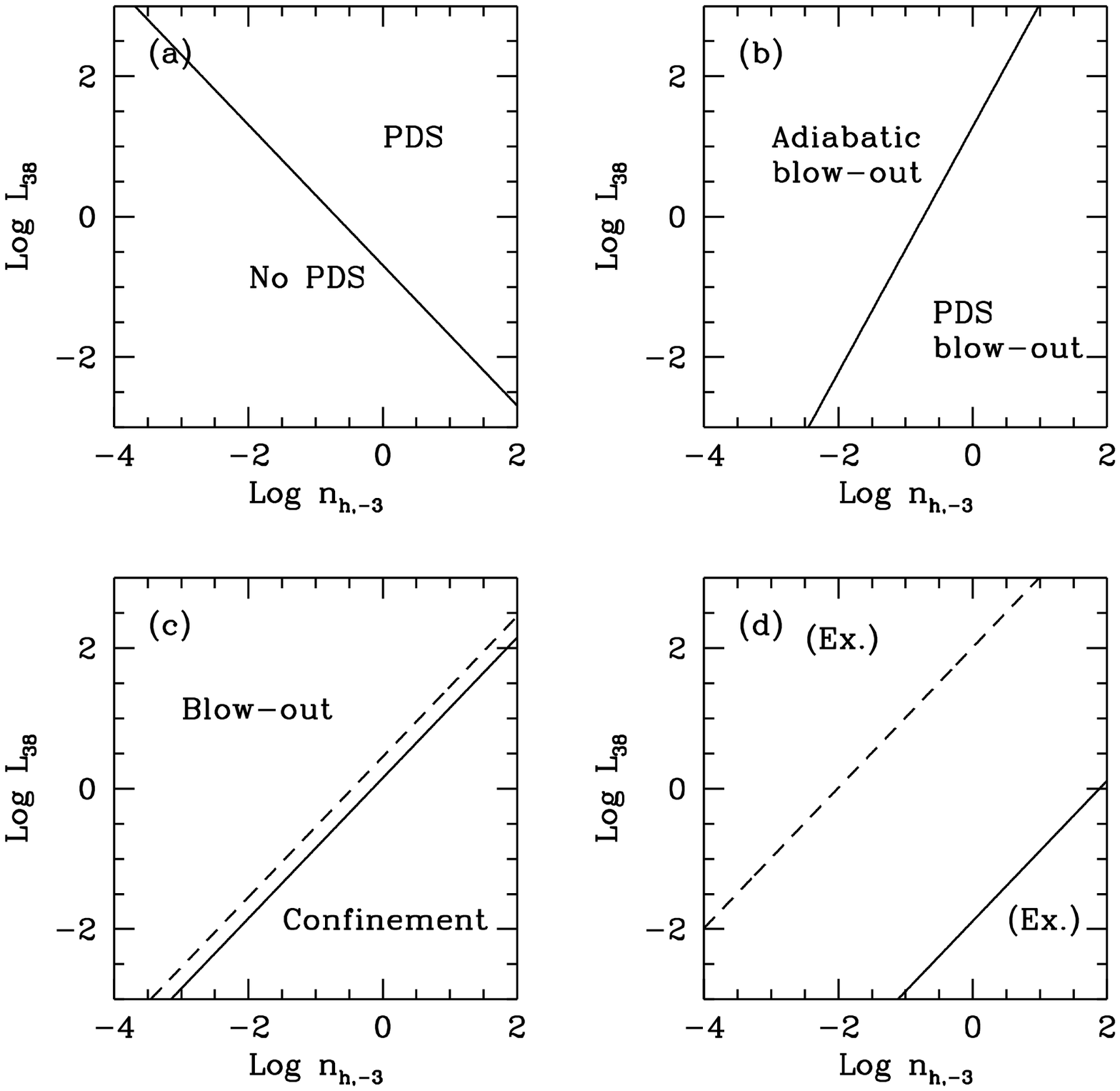}
}
\caption{Fate of SBs in the \nh--\lmech\ plane.  The panels indicate
the lines separating the domains in which SBs (a) get or do not get
into the PDS stage, (b) blow-out in the adiabatic or PDS stage, (c)
blow out or remain confined, (d) end after exhaustion of SNe.}
\label{fig:plane1}
\end{figure}

It is useful to analyse the fate of SBs in the \nh--\lmech\ plane.
The condition $\Tpds = \Tconfad$ determines the possibility of getting
into the PDS stage; it is equivalent to the line:

\be L_{38} = 0.20\ T_{\rm h,6}^{11/2}\ n_{\rm h,-3}^{-1}\, .
\label{eq:condpds} \ee

\noindent
In absence of blow-out, below this line SBs are pressure-confined in
the adiabatic stage, above this line they can go into the PDS stage
(Fig.~\ref{fig:plane1}a).  The condition $\Tpds = \Tbos$ determines
whether the blow-out is in the adiabatic or PDS stage; it is
equivalent to:

\be L_{38} = 19\ H_{\rm eff,3}^{11/4}\ n_{\rm h,-3}^{7/4}\, .
\label{eq:condboad} \ee

\noindent
In case of blow-out, to the left of this line SBs blow out in the
adiabatic stage (which is maintained even after one sound crossing
time), to the right of it SBs blow out in the PDS stage
(Fig.~\ref{fig:plane1}b).  The condition $\Tbo=\Tconf$ determines
whether SBs are going to end by blow-out or confinement; in the
adiabatic stage it is equivalent to:

\be L_{38} = 1.4 \ H_{\rm eff,3}^2\ T_{\rm h,6}^{3/2}\ 
n_{\rm h,-3}\, .
\label{eq:condbo} \ee

\noindent
In the PDS stage the numerical factor is 2.9.  Below this line SBs are
pressure-confined, above it they blow out (Fig.~\ref{fig:plane1}c).

\begin{table}
\begin{center}
\begin{tabular}{lcll}
\hline
\tpds &=& $2.87\times10^6\ L_{38}^{3/11} n_{\rm h,-3}^{-8/11} \mu_{\rm
h,0.6}^{9/22} \zeta_{\rm m}^{-5/11}$ yr\\
\rpds &=& $674\ L_{38}^{4/11} n_{\rm h,-3}^{-7/11} \mu_{\rm h,0.6}^{1/22}
\zeta_{\rm m}^{-3/11}$ pc \\ 
&&\\
\tconf &=& $4.12\times10^6\ L_{38}^{1/2} n_{\rm h,-3}^{-1/2} \mu_{\rm h,0.6}^{3/4}
T_{\rm h,6}^{-5/4}$ yr &(ad.)\\
       &=& $2.95\times10^6\ L_{38}^{1/2} n_{\rm h,-3}^{-1/2}
\mu_{\rm h,0.6}^{3/4} T_{\rm h,6}^{-5/4}$ yr &(PDS)\\
\rconf &=& $857\ L_{38}^{1/2} n_{\rm h,-3}^{-1/2} \mu_{\rm h,0.6}^{1/4} 
T_{\rm h,6}^{-3/4}$ pc &(ad.)\\
       &=& $592\ L_{38}^{1/2} n_{\rm h,-3}^{-1/2} \mu_{\rm h,0.6}^{1/4} 
T_{\rm h,6}^{-3/4}$ pc &(PDS)\\
&&\\
\tbo   &=& $5.53\times10^6\ L_{38}^{-1/3} n_{\rm h,-3}^{1/3}
\mu_{\rm h,0.6}^{1/3} H_{\rm eff,3}^{5/3}$ yr &(ad.)\\
       &=& $7.06\times10^6\ L_{38}^{-1/3} n_{\rm h,-3}^{1/3}
\mu_{\rm h,0.6}^{1/3} H_{\rm eff,3}^{5/3}$ yr &(PDS)\\
$R_{\rm bo}$ &=& $10^3\ H_{\rm eff,3}$ pc \\
\tbos  &=& 3.11 \tbo & (ad)\\
       &=& 2.81 \tbo & (PDS)\\
\rbos  &=& 1.98 $R_{\rm bo}$ & (ad)\\
       &=& 1.86 $R_{\rm bo}$ & (PDS)\\
\hline
\end{tabular}
\label{table:sb2}
\caption{Characteristic times of a SB for typical values of the
parameters.  Here $\mu_{\rm h,0.6}=\Muh/0.6$.}
\end{center}\end{table}

\begin{figure}
\centerline{
\includegraphics[width=9cm]{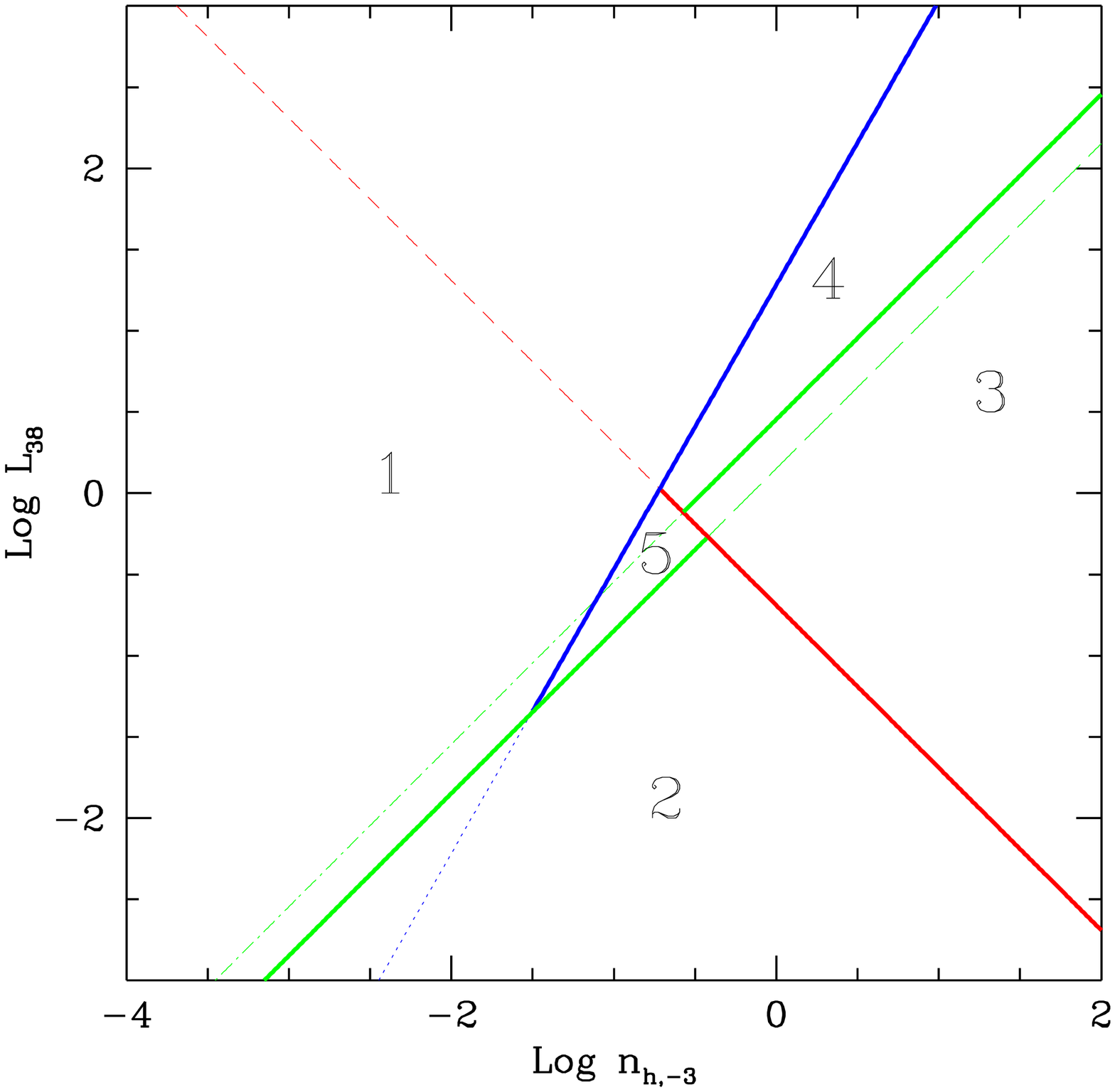}
}
\caption{As in Fig.~\ref{fig:plane1}, but with all lines drawn together.
For the explanations of the five regions, see the text.}
\label{fig:plane2}
\end{figure}

Fig.~\ref{fig:plane2} shows all the relations listed above and the
regions they define in the \nh--\lmech\ plane for the choice
$\Th=10^6\ K$ and $\Heff=10^3$ pc.  The thick lines mark the
boundaries of regions where SBs end in a different way.  Region (1) in
the figure contains the SBs that blow out in the adiabatic stage.  In
region (2) SBs are confined in the adiabatic stage, while in region
(3) SBs are confined in the PDS stage.  In region (4) SBs blow out in
the PDS stage.  Finally, in the closed region (5) SBs blow out in the
adiabatic stage, but are confined before one sound crossing time.  For
other values of \heff\ and \th\ the lines move along the plane, but it
is easy to check that the area of the triangle corresponding to region
(5) in invariant.

It is interesting also to consider the region of the parameter space
in which SBs end after all SNe have exploded.  This happens when
$t_{\rm bo} = \Tlife$ and $t_{\rm conf} = \Tlife$.  The two conditions
give:

\be \begin{array}{ccl}
L_{38} &=& 6.2\times10^{-3}\, H_{\rm eff,3}^5   n_{\rm h,-3}\\
L_{38} &=& 1.9\times10^{2} \, T_{\rm h,6}^{5/2} n_{\rm h,-3}\, .\\
\end{array} \label{eq:condex} \ee

\noindent
They are both satisfied below the first line and above the second
(Fig.~\ref{fig:plane1}d).  As the two lines are parallel, this
happens only if $H_{\rm eff,3}T_{\rm h,6}^{-1/2}=7.9$.  In this case a
whole band in the plane corresponds to exhaustion.

We conclude this discussion by noting that the plane is not uniformly
populated, as both \nh\ and \lmech\ are dynamical variables.  In other
words, the probability of a certain regime is not determined by the
area occupied in the plane but by the dynamics of the system.

\bsp

\label{lastpage}

\end{document}